\documentclass[default,iicol,sn-mathphys]{sn-jnl}%

\jyear{2021}%

\theoremstyle{thmstyleone}%
\theoremstyle{thmstyletwo}%

\theoremstyle{thmstylethree}%

\newcommand{\pore}{{\mathrm{pore}}}
\newcommand{\FF}{{\mathrm{ff}}}
\newcommand{\PM}{{\mathrm{pm}}}
\newcommand{\BJ}{{\mathrm{BJ}}}
\newcommand{\mm}{{\mathrm{mm}}}
\newcommand{\m}{{\mathrm{m}}}
\renewcommand{\vec}[1]{{\ensuremath{\boldsymbol{\mathrm #1}}}}
\newcommand{\ten}[1]{\ensuremath{\boldsymbol{\mathsf{#1}}}}
\newcommand{\vdot}{\boldsymbol{\mathsf{\ensuremath\cdot}}}
\newcommand{\del}{\ensuremath{\nabla}}
\newcommand{\deld}{\ensuremath{\del\vdot}}

\begin{document}

\title[A Surrogate-Assisted Uncertainty-Aware Bayesian Validation Framework]{A Surrogate-Assisted Uncertainty-Aware Bayesian Validation Framework and its Application to Coupling Free Flow and Porous-Medium Flow}


\author*[1]{Farid Mohammadi}\email{farid.mohammadi@iws.uni-stuttgart.de}
\author[2]{Elissa Eggenweiler}\email{elissa.eggenweiler@ians.uni-stuttgart.de}
\author[1]{Bernd Flemisch}\email{bernd.flemisch@iws.uni-stuttgart.de}
\author[1]{Sergey Oladyshkin}\email{sergey.oladyshkin@iws.uni-stuttgart.de}
\author[2]{Iryna Rybak}\email{iryna.rybak@ians.uni-stuttgart.de}
\author[1]{Martin Schneider}\email{martin.schneider@iws.uni-stuttgart.de}
\author[1]{Kilian Weishaupt}\email{kilian.weishaupt@iws.uni-stuttgart.de}

\affil*[1]{\orgdiv{Institute for Modelling Hydraulic and Environmental Systems}, \orgname{University of Stuttgart}, \orgaddress{\street{Pfaffenwaldring 61}, \city{Stuttgart}, \postcode{70569}, \country{Germany}}}

\affil[2]{\orgdiv{Institute of Applied Analysis and Numerical Simulation}, \orgname{University of Stuttgart}, \orgaddress{\street{Pfaffenwaldring 57}, \city{Stuttgart}, \postcode{70569}, \country{Germany}}}

\abstract{
Existing model validation studies in geoscience often disregard or partly account for uncertainties in observations, model choices, and input parameters.
In this work, we develop a statistical framework that incorporates a probabilistic modeling technique using a fully Bayesian approach to perform a quantitative uncertainty-aware validation. A Bayesian perspective on a validation task yields an optimal bias-variance trade-off against the reference data. It provides an integrative metric for model validation that incorporates parameter and conceptual uncertainty.
Additionally, a surrogate modeling technique, namely Bayesian Sparse Polynomial Chaos Expansion, is employed to accelerate the computationally demanding Bayesian calibration and validation.
We apply this validation framework to perform a comparative evaluation of models for coupling a free flow with a porous-medium flow.
The correct choice of interface conditions and proper model parameters for such coupled flow systems is crucial for physically consistent modeling and accurate numerical simulations of applications.
We develop a benchmark scenario that uses the Stokes equations to describe the free flow and considers different models for the porous-medium compartment and the coupling at the fluid--porous interface.
These models include a porous-medium model using Darcy's law at the representative elementary volume scale with classical or generalized interface conditions and a pore-network model with its related coupling approach.
We study the coupled flow problems' behaviors considering a benchmark case, where a pore-scale resolved model provides the reference solution.
With the suggested framework, we perform sensitivity analysis, quantify the parametric uncertainties, demonstrate each model's predictive capabilities, and make a probabilistic model comparison.}

\keywords{Model validation, uncertainty quantification, free flow, porous-medium flow, interface conditions}


\pacs[MSC Classification]{65C20, 62F15, 65C05, 35Q35, 76D07, 76M12, 76S05}

\maketitle

\section{Introduction}
\label{sec:intro}
Geoscientists use modeling as an essential instrument to study the causes and effects of their problems, with an increasing level of detail and complexity.
Over the last century, computational modeling in geoscience, especially in porous media research, has witnessed tremendous improvement. 
After decades of development, the state-of-the-art simulators can now solve coupled partial differential equations governing the complex subsurface multiphase flow behavior within a practically large spatial and temporal domain.
Given the importance of computational modeling, assessing the models' reliability is paramount to engineering designers, managers, public officials, and those affected by the decisions based on the predictions.
For this reliability assessment, model validation is commonly used, among other complementary measures, to assess the model performance.
The relevant hypotheses regarding assessing the reliability of a model could be building up or losing trust in its generated simulations. In terms of validation, the hypothesis is whether the model can satisfactorily represent the real system of interest.

The word ``validation'' is commonly used to refer to simple comparisons between model outputs and experimental data or any reference data. Usually, these comparisons constitute plotting the model results against data on the same axes to assess agreement or lack thereof visually.
However, they are clearly insufficient as a basis for making decisions regarding model validity. 
These naive comparisons often disregard or only partly account for existing uncertainties in the experimental observations or the model input parameters. It is impossible to appropriately determine whether the model and data agree without accounting for these uncertainties.

Moreover, for assessing the physical phenomena in question, several representations, i.e., models, might exist with different approaches and assumptions to analyze the occurring processes. 
Therefore, a significant research challenge is accurately assessing competing modeling concepts and validating the corresponding computational models against an experiment or a reference solution.
In the case of the multimodel comparison, the validation hypothesis is which model within the pool of available models can represent the reality, i.e., observed values in the experiments or the reference data resulting from a detailed simulation.
This may be analyzed as a validation benchmark.
Following the guidelines for validation benchmarks in \cite{oberkampf2010verification,OberkampfEtAl:2008:VVB}, the most important scientific question to be addressed here is how to compare the computational models quantitatively. 
Each validation benchmark must declare one or more methods that should be employed for quantitative comparison. 
The development of such validation metrics is an active field of research. While being potentially applicable to many research areas, rigorously developed validation metrics have been applied to some of them only, such as computational fluid dynamics \cite{oberkampf2006measures}, computational solid mechanics \cite{mahadevan2005validation} or heat transfer \cite{hills2006model}.

To address the concerns above in the model validation, we developed an uncertainty-aware validation framework that includes a rigorous uncertainty quantification of models and provides validation metrics for comparing the competing models.
This framework employs the Bayesian probability theory, which offers a robust framework for characterizing scientific inference since its simple concept lies in the fact that rational belief comes in degrees that can be measured in terms of probabilities.
Moreover, a Bayesian perspective on a validation task yields an optimal bias-variance trade-off against the experimental data or reference solution. It provides an integrative metric for model validation that incorporates parameter and conceptual uncertainty~\cite{geman1992neural,schoniger2014model,schoniger2015statistical}.
Thus, it is proven to form a viable method for data-driven validation of computer simulations and can provide a solid basis for a sound evaluation of computer simulations.
Following the Bayesian model selection pathway ~\cite{draper1995,hoeting1999bayesian}, we employ the Bayesian model evidence (BME) as a validation score, indicating the quality of the analyzed physical models against the available experimental data or finely resolved reference solution.

The probabilistic nature of the Bayesian technique requires propagating the parametric uncertainty through all competing models -- i.e., a significant number of model evaluations -- to reach statistical convergence. In practice, however, the computational complexity of the underlying computational model and the total available computational budget severely restrict the number of evaluations one can perform. This challenge renders the brute-force computation of the BME value, required for the Bayesian validation framework (BVF), infeasible.
We replace each original computational model with its easy-to-evaluate surrogate in the Bayesian analysis.
A surrogate-assisted Bayesian analysis has been applied to many applications, including hydrology, e.g.,~\cite{yoon2022bayesian}, sediment transport, e.g.,~\cite{mohammadi2018bayesian, beckers2020bayesian}, processes in subsurface reservoirs, e.g.,~\cite{bazargan2015surrogate, bazargan2017bayesian, scheurer2021surrogate}, subsurface flow models, e.g.,~\cite{elsheikh2014efficient}.
A surrogate model's primary goal is to replicate the behavior of the underlying physical models from a limited set of runs without sacrificing accuracy.
 
For constructing a surrogate, the computational model needs to be evaluated using those sets of modeling parameters, out of various possibilities, that cover the parametric space as well as possible. The polynomial chaos expansion (PCE)~\cite{Wiener1938} or its extension toward arbitrary polynomial chaos expansion (aPCE) \cite{oladyshkin2012data} is a well-known and rigorous approach to providing functional representations of stochastic quantities. 
However, not all expansion terms are relevant for representing the underlying physical processes, and employing the concept of sparsity can lead to zero values for many coefficients in the expansion~\cite{tipping2001sparse}. 
Many mathematical approaches exist when dealing with a regression problem, such as PCE representation, that lead to a sparse solution. These approaches have led to the emergence of numerous sparse solvers in the compressed sensing \cite{arjoune2017compressive}, as well as in the sparse PCE. In~\cite{lthen2020sparse} a comprehensive survey of the proposed solvers in the context of PCE is provided.
Here, we employ Bayesian sparse aPCE denoted as BsaPCE, which is an extension of aPCE within a Bayesian framework using a Bayesian sparse learning method \cite{tipping2003fast}. BsaPCE not only identifies the expansion terms, which capture the physical model's relevant features, but it can also provide a probabilistic prediction, i.e., a prediction with the associated uncertainty. This prediction uncertainty can be used as the expected surrogate uncertainty when replacing the original computational model with a possibly less accurate surrogate.

In this study, we apply the proposed surrogate-assisted Bayesian validation framework to coupling free flow and porous-medium flow. 
Coupled free-flow and porous-medium systems play a significant role in many industrial, environmental, and biological settings, e.g., fuel cells, water flows in karst aquifers, blood flows in vessels, and living tissues. Flow interaction between the free-flow region\nocite{*} and the porous-medium domain is highly involved and strongly interface-driven. Therefore, a physically consistent description of flow processes in the whole coupled system, especially near the interface, is crucial for accurate numerical simulations of applications. 
A lot of effort has been made during the last decades in mathematical modeling and analysis of such coupled flow systems, and several coupling strategies have been proposed, e.g.,~\cite{Angot_etal_17, Discacciati_Quarteroni_09, Goyeau_Lhuillier_etal_03,  Jaeger_Mikelic_09, Lacis_Bagheri_17, OchoaTapia_Whitaker_95,Eggenweiler_Rybak_MMS20}.
The possibilities to conceptualize these coupling conditions could be regarded as conceptual uncertainty. This conceptual uncertainty is mainly related to the description of processes in the porous medium and near the interface, for which, different mathematical models and coupling strategies are considered.

We use the literature's most widely studied coupled flow problem, namely the Stokes--Darcy problem. In this setting, the free-flow conceptualization is based on the Stokes equations for all discussed models. However, the way these models simulate the fluid flow in the porous medium and the set of coupling conditions imposed on the fluid--porous interface varies. 
Apart from the conceptual uncertainty, each computational model contains parametric uncertainty, such as material parameters or interface location that must also be rigorously addressed.
To the best of our knowledge, no rigorous uncertainty-aware validation for this application has been performed so far.

The rest of this paper is organized as follows.
In Section~\ref{sec:bayes}, we introduce the Bayesian validation framework and discuss how it can be accelerated by means of surrogate modeling.
Section~\ref{sec:models} presents the mathematical and computational models that are going to be evaluated by means of the proposed benchmark scenario.
The benchmark scenario is described in Section~\ref{sec:scenarios}, while Section~\ref{sec:results} is devoted to a discussion of the application of the Bayesian model selection.
In Section~\ref{sec:conclusions}, we present a summary and conclusion to our research.

\section{Bayesian validation framework}
\label{sec:bayes}
The current section introduces the BVF for comparing and validating computational models. The performance of these models is compared to a reference, either observed data from experiments or highly-resolved reference models.
The benefits of this comparison are twofold. First, this evaluates the strengths and weaknesses of competing modeling concepts. Second, the predictive ability of each computational model is assessed. 

Furthermore, we update the prior belief in the predictive capability of the models based on Bayesian notions. The resulting so-called posterior belief is expressed in terms of probabilities. The Bayesian approach in the validation task allows us to include possible sources of errors that can lead to inevitable uncertainties. The BVF requires propagation of the parametric uncertainty through the given computationally demanding models. This propagation renders the analysis intractable, as it demands many model evaluations. To circumvent this problem, we employ a surrogate modeling technique to offset BVF's computational time.

\subsection{Bayesian model comparison}\label{sec:BayesModelComparison}
The topic of quantitative model comparison has received and continues to receive considerable attention in the field of statistics. There exist several multimodel comparison frameworks related to these model rating methods that allow for statistical model selection and averaging, e.g., \cite{gelman2013bayesian}.
The most common approach is Bayesian model selection (BMS) \cite{draper1995,hoeting1999bayesian}. 
BMS is grounded on Bayes’ theorem, which combines a prior belief about the efficiency of each model with its performance in replicating a common observation data set. Its procedure for model comparison entails principled and general solutions to the trade-off between parsimony and goodness-of-fit.
Moreover, BMS is a formal statistical approach that allows comparing alternative conceptual models, testing their adequacy, combining their predictions into a more robust output estimate, and quantifying conceptual uncertainty's contribution to the overall prediction uncertainty. BMS can be regarded as a Bayesian hypothesis testing framework, combining the idea of classical hypothesis testing with the ability to examine multiple alternative models against each other in a probabilistic manner. It returns the so-called model weights~\cite{geman1992neural} representing posterior probabilities for each model to be the most appropriate from the set of proposed competing models. Thus, the computed model weights provide a quantitative ranking for the competing conceptual models.

Let us consider $N_m$ competing computational models $M_k$, each with an uncertain parameter vector $\mathbf{\theta}_k$ of length $N_k$, yielding a quantity $Q$ of interest  in a physical space of $x$, $y$ and $z$ and for a time stamp of $t$.
The model weights are given by Bayes’ theorem, which can be cast for a set of $M_k$ competing models as
\begin{equation}
\label{eq:BayesBMA}
    P(M_k\vert\mathcal{Y}) = \frac{p(\mathcal{Y} \vert M_k) P(M_k)}{\sum_{i=1}^{N_m} p(\mathcal{Y}\vert M_i) P(M_i)},
\end{equation}
%
where $P(M_k)$ denotes the prior probability of the model, also known as the subjective credibility that model $M_k$ could be the most plausible model in the set of models \textit{before} any comparison with observed data have been made.
The term $p(\mathcal{Y}\vert M_k)$ is the Bayesian model evidence (BME), also known as marginal likelihood, of the model $M_k$.  Bayes’ theorem closely follows the principle of parsimony or Occam’s razor \cite{angluin1983inductive}, in that the posterior model weights $P(M_k \vert\mathcal{Y})$ offer a compromise between model complexity and goodness of fit, known as the bias-variance trade-off \cite{geman1992neural}. The model weights, $P(M_k \vert \mathcal{Y})$, can be interpreted as the Bayesian probability of the individual models to be the best representation of the system from the pool of competing models.

Hoeting et al.~\cite{hoeting1999bayesian} proposed that a `reasonable, neutral choice' could be equally likely priors, i.e., $P(M_k)=1/N_m$, in case of paucity of prior knowledge regarding the merit of the different models under consideration. The denominator in~\eqref{eq:BayesBMA} is the normalizing constant of the posterior distribution of the models and can simply be obtained by determination of the individual weights. Since all model weights are normalized by the same constant, this normalizing factor could even be neglected. Thus, the weights $P(M_k \vert \mathcal{Y})$ of the individual model $M_k$ against other models can be represented by the proportionality
\begin{equation}
\label{eq:Proportionality}
    P(M_k\vert \mathcal{Y}) \propto p(\mathcal{Y}\vert M_k) P(M_k).
\end{equation}
%

The BME term $p(\mathcal{Y}\vert M_k)$ quantifies the likelihood of the observed data based on the prior distribution of the parameters. It can be computed by integrating the likelihood term in Bayesian theorem \cite{kass1995bayes} over the parameter space $\Theta_k$ of the model $M_k$:
\begin{equation}
\label{eq:BME}
    p(\mathcal{Y}\vert M_k) = \int_{\Theta_k} p(\mathcal{Y}\vert M_k,\mathbf{\theta}_k) P(\mathbf{\theta}_k\vert M_k) d\mathbf{\theta}_k,
\end{equation}
%
with $\mathbf{\theta}_k$ being the parameter vector from the parameter space $\Theta_k$ of model $M_k$. The term $P(\mathbf{\theta}_k \vert M_k)$ is the corresponding prior distribution of parameters $\mathbf{\theta}_k$ for the model $M_k$. Assuming that the measurement errors follow a Gaussian distribution, the likelihood or probability of the parameter set $\mathbf{\theta}_k$ of model $M_k$ to have generated the observation data with the independent realization of $\mathcal{Y}=\left(\mathcal{Y}_1, ..., \mathcal{Y}_N\right)^{\top}$ is represented by
\begin{equation}
\label{eq:BayesLikelihood}
    \begin{split}
    p&(\mathcal{Y} \vert M_k,\mathbf{\theta}_k) := \prod_{i=1}^{N}p(\mathcal{Y}_i \vert M_k,\mathbf{\theta}_k)\\
    =&\frac{1}{\sqrt{(2 \pi)^{N} \operatorname{det} \Sigma}} \times \\ 
    &\exp \left(\!\!-\frac{(M_k(\theta_k)\!-\!\mathcal{Y})^{T} \Sigma^{-1}(M_k(\theta_k)\!-\!\mathcal{Y})}{2}\!\!\right),
    \end{split}
\end{equation}
%
where $\Sigma$ denotes the covariance matrix, which includes all error sources to be explained in Section~\ref{sec:square-UQ}. To obtain the BME values for each competing model, we estimate the integral in \eqref{eq:BME} using a brute-force Monte Carlo integration~\cite{Smith1992} avoiding unnecessary assumptions. For more details on the properties of BME and a comparison of various techniques to estimate this term, the reader is referred to~\cite{schoniger2014model} and~\cite{Oladyshkin_2019}.

The ratio of BME values for two alternative models is defined as the Bayes Factor $BF(M_k , M_l)$%
, which is a key component in Bayesian hypothesis testing framework introduced in~\cite{jeffreys1961theory}:
\begin{equation}
\begin{split}
\label{eq:BayesFactor}
    BF(M_k , M_l) &= \frac{P(M_k \vert \mathcal{Y})}{P(M_l \vert \mathcal{Y})} \frac{P(M_l)}{P(M_k)} \\
    &= \frac{p(\mathcal{Y} \vert M_k)}{p(\mathcal{Y}\vert M_l)}.
\end{split}
\end{equation}
It quantifies the evidence (literally, as in Bayesian model evidence) of the hypothesis $M_k$ against the null-hypothesis $M_l$. Stated differently, the Bayes Factor $BF(M_k , M_l)$, can be interpreted as the ratio between the posterior and prior odds of model $M_k$ being the more plausible one in comparison to the alternative model $M_l$ \cite{kass1995bayes}. Jeffreys~\cite{jeffreys1961theory} provides a rule of thumb for the interpretation of Bayes Factor. The grades of evidence are summarized in Table~\ref{tab:BayesFactorThresholds}. Following this suggestion, a Bayes Factor which lies between 1 and 3 indicates an evidence in favor of $M_k$ that is `not worth more than a bare mention', a factor of up to 10 represents `substantial' evidence, and a factor between 10 and 100 can be regarded a `strong' evidence. Finally, a Bayes Factor greater than 100 admits `decisive' evidence, i.e., it can be used as a threshold to reject models based on poor performance in comparison to the best performing model in the set of considered models.

\begin{table}[!t]
\caption{Interpretation of Bayes Factor in favor of model $M_k$ according to \cite{jeffreys1961theory}.}
\label{tab:BayesFactorThresholds}
\centering
\begin{tabular}{l l}
\hline
Bayes Factor ($BF$) & Interpretation  \\ 
\hline
1 -- 3  & anecdotal evidence \\
3 -- 10  & substantial evidence \\
10 -- 100  & strong evidence \\
$>$ 100  & decisive evidence \\
\hline
\end{tabular}
\end{table}
%

\subsection{Accelerating the analysis via surrogate modeling}
\label{sec:Surrogate}
As stated earlier, BVF requires uncertainty propagation through each competing model, demanding a significant number of model evaluations to yield statistical convergence. In practice, however, the computational complexity of the underlying computational models and the total available computational budget severely restrict the number of evaluations one can carry out. In such situations, the Bayesian analysis estimations lack sufficient trust, as the limited number of model evaluations can yield additional uncertainty.
To tackle this challenge, we replace each competing model's response in the Bayesian validation framework with an easy-to-evaluate surrogate representation using the theory of PCE introduced in~\cite{Wiener1938}.

The PCE representation of the model $M_k$ provides the dependence of the computational model $M_k$ on the uncertain model's parameters $\mathbf{\theta}_k$ using projection onto an orthonormal polynomial basis~\cite{oladyshkin2018incomplete}. It could be also seen as a linear regression that includes linear combinations of a fixed set of nonlinear functions with respect to the input variables, known as polynomial basis function
\begin{equation}
\label{eq:PCE_Trunc}
    M_k(x,y,z,t, \mathbf{\theta}_k) \!\approx \!\sum_{\vec{\alpha} \in \mathcal{A} } c_{\vec{\alpha}} (x,y,z,t) \Psi_{\vec{\alpha}}(\mathbf{\theta}_k).
\end{equation}
Here, $x,y,z,t$ are the spatial and temporal components of the quantity of interest, $\mathbf{\theta}_k$ is the vector of the $N_k$ uncertain parameters of model $M_k$, $c_{\vec{\alpha}}(x,y,z,t)  \in \mathbb{R}$ are the corresponding expansion coefficients that are functions of space and time, and $\Psi_{\vec{\alpha}}(\mathbf{\theta}_k)$ denotes multivariate polynomials orthogonal with respect to a multi-index~$\vec{\alpha}$. The latter represents the combinatoric information how to enumerate all possible products of $N_k$ individual univariate basis functions with respect to the total degree of expansions less or equal to polynomial degree $d$ \cite{marelli2015uqlab}:
\begin{equation}
\label{eq:truncation}
\begin{split}
    \mathcal{A}^{N_k, d} = \{ \vec{\alpha} \in \mathbb{N}^{N_k} \ : \ \lvert\vec{\alpha}\rvert\leq d\} \, , \\
    \operatorname{card} \ \mathcal{A}^{N_k, d} \equiv P = \binom{N_k+d}{d}.
\end{split}
\end{equation}

The multivariate polynomials $\Psi_{\vec{\alpha}}(\mathbf{\theta}_k)$ are comprised of the tensor product of univariate polynomials
\begin{equation}
\label{eq:Psi}
    \Psi_{\vec{\alpha}}(\mathbf{\theta}_k) :=  \prod_{i=1}^{N_k} \psi_{\alpha_i}^{(i)}(\mathbf{\theta}_{k,i}) \, ,
\end{equation}
where the univariate orthonormal polynomials $\psi_{\alpha_i}^{(i)}(\mathbf{\theta}_{k,i})$ must satisfy 
\begin{equation}
\begin{split}
\label{eq:univPsi}
    &\langle \psi_j^{(i)}(\mathbf{\theta}_{k,i}), \psi_l^{(i)}(\mathbf{\theta}_{k,i}) \rangle := \\ &\int_{\Theta_{k,i}} \! \! \! \!  \psi_j^{(i)}(\mathbf{\theta}_{k,i}) \psi_l^{(i)}(\mathbf{\theta}_{k,i}) f_{\Theta_{k,i}}  (\mathbf{\theta}_{k,i})d \mathbf{\theta}_{k,i} \! = \delta_{j l} \, .
\end{split}
\end{equation}

Here, $i$ represents the input variable with respect to which the polynomials are orthogonal as well as the corresponding polynomial family, $j$ and $l$ are the corresponding polynomial degree, $f_{\Theta_{k,i}}(\mathbf{\theta}_{k,i})$ is the $i$th-input marginal distribution and $\delta_{j l}$ is the Kronecker delta.
We use an arbitrary version of PCE, namely aPCE, introduced in~\cite{oladyshkin2012data}, that can operate with probability measures that may be implicitly and incompletely defined via their statistical moments. Using aPCE, one can build the multivariate orthonormal polynomials even in the absence of the exact probability density function $f_{\Theta_k}(\theta)$.

The main task of the surrogate representation of model $M_k(x,y,z,t, \mathbf{\theta}_k)$ in~\eqref{eq:PCE_Trunc} is to compute the coefficients $c_{\vec{\alpha}}$. However, not all coefficients could be relevant for such a surrogate. Therefore, we employ the Bayesian sparse learning method~\cite{tipping2001sparse} using a fast marginal likelihood maximization algorithm~\cite{tipping2003fast}. Doing so, we sequentially identify the relevant predictors (expansion terms) that capture the most significant features of the physical model. We denote this extension of aPCE as Bayesian sparse arbitrary polynomial chaos (BsaPCE) representation \cite{mohammadi2020development}. The posterior distribution of the expansion coefficients, conditioned on the model responses $\mathrm{\mathbf{Y}}$ resulting from the training sets $\mathbf{X}$, is given by the combination of a Gaussian likelihood and a Gaussian prior distribution over the unknown expansion coefficients $\mathbf{c}$ according to Bayes' rule. Then, the posterior of the expansion coefficients given the model responses $\mathrm{\mathbf{Y}}$ and values of hyper-parameters $\vec{\alpha}$ and $\beta$ describing the Gauss process \cite{oladyshkin2020bayesian3}, can take the following form
\begin{equation}
\label{eq:PCE_Posterior}
    p(\mathrm{\mathbf{c}}\vert\mathbf{Y},\vec{\alpha}, \beta) = \frac{p(\mathrm{\mathbf{Y}}\vert \mathbf{X},\mathbf{c}, \beta) p(\mathbf{c}\vert\vec{\alpha})}{p(\mathrm{\mathbf{Y}}\vert \mathbf{X}, \vec{\alpha} , \beta)},
\end{equation}
which is also Gaussian defined by $\mathcal{N}( \mathbf{c} \vert \vec{\mu}, \mathbf{\Sigma})$ with
\begin{equation}
\label{eq:PCE_Posterior_moments}
    \vec{\mu} = \beta \mathbf{\Sigma} \mathbf{\Psi}^{\top} \mathrm{\mathbf{Y}} \, , \qquad
    \mathbf{\Sigma} = \left(\mathbf{A}+ \mathbf{\Psi}^{\top} \beta \mathbf{\Psi} \right)^{-1} \, .
\end{equation}
%
Here, $\mathbf{\Psi}$ is the design matrix of size $E \times N$ with elements $\Psi_{ni}=\psi_i(x_n)$, where $E$ represents the number of model evaluations using the training samples, and $\mathbf{A}=\mathrm{diag}(\alpha_i)$. The values of $\vec{\alpha}$ and $\beta$ can be determined via type-II maximum likelihood \cite{berger2013statistical}.
Having found values $\vec{\alpha}^*$ and $\mathbf{\beta}^*$ for the hyperparameters that maximize the marginal likelihood, one can evaluate the predictive distribution over $\mathrm{Y}$ for a new input $\mathrm{\mathbf{x}}$ by
\begin{equation}
\label{eq:Pred_Dist}
\begin{split}
    p(\mathrm{Y} & \vert \mathbf{x}, \mathbf{X}, \mathrm{\mathbf{Y}}, \vec{\alpha}^* , \beta^*) 
    \\
    &= \int p(\mathrm{Y}\vert \mathbf{x}, \mathbf{c} , \beta^*) p(\mathbf{c}\vert\mathbf{X}, \mathrm{\mathbf{Y}}, \vec{\alpha}^*, \beta^*) d\mathbf{c} \\
    &= \mathcal{N}(\mathrm{Y} \vert \mathbf{\mu}^\top \Psi(\mathbf{x}), \sigma^2(\mathbf{x})) \, .
\end{split}
\end{equation}

The predictive mean is given by Equation~\eqref{eq:PCE_Posterior_moments} with $\mathbf{c}$ set to the posterior mean $\vec{\mu}$, and the variance of the predictive distribution is given by
\begin{equation}
\label{eq:Pred_Dist_sigma2}
    \sigma^2(\mathbf{x}) = (\beta^*)^{-1} + \Psi(\mathbf{x})^\top \mathbf{\Sigma} \Psi(\mathbf{x}) \, ,
\end{equation}
%
where $\mathbf{\Sigma}$ is calculated by Equation~\eqref{eq:PCE_Posterior_moments} in which $\vec{\alpha}$ and $\beta$ set to their optimized values $\vec{\alpha}^*$ and $\beta^*$.  For the latter, a separate hyperparameter $\alpha_i$ is assigned to each weight parameter $c_i$. 

\section{Mathematical and computational models}
\label{sec:models}

In this section, we provide the assumptions on the  coupled flow system of interest in his work and introduce the flow models for which the Bayesian validation framework is applied in Section~\ref{sec:results}.

\begin{figure}[ht]
    \includegraphics[scale=0.84]{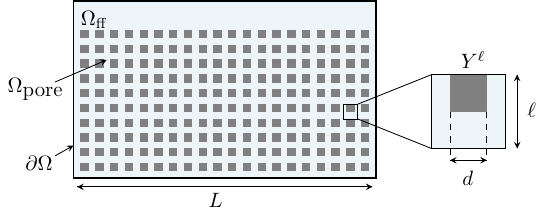} \\ \hspace*{4ex}
    \quad 
    \includegraphics[scale=0.86]{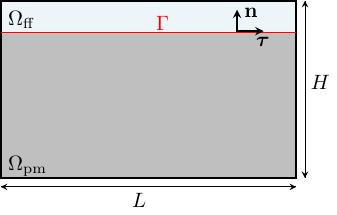}
    \caption{Geometrical setting at the pore scale (top) and REV scale (bottom).}
    \label{fig:setting}
\end{figure}

From a pore-scale perspective, we consider a two-dimensional flow domain $\Omega_\text{flow}$ 
consisting of the free-flow domain $\Omega_\FF$ and the pore space~$\Omega_\text{pore}$ of the porous medium. The porous-medium domain~$\Omega_\PM$ has a periodic structure composed by the repetition of the representative elementary volume (REV) (scaled unit cell) 
$Y^\ell=(0,\ell) \times (0,\ell)$, where $\ell$ 
is the microscopic length scale (Figure~\ref{fig:setting}, top). From a macroscopic point of view, the coupled domain $\Omega = \Omega_\FF \cup \Omega_\PM$ comprises the free-flow region $\Omega_\FF$ and the porous-medium domain $\Omega_\PM$, separated by a sharp fluid--porous interface $\Gamma$ (Figure~\ref{fig:setting}, bottom).
\enlargethispage{0.5cm}

\vspace*{-0.1cm}
We consider isothermal single-phase steady-state flow at low Reynolds numbers. The same fluid occupies the free-flow domain and fully saturates the porous medium. This fluid is supposed to be incompressible and to have constant viscosity. We consider a non-deformable porous medium leading to constant porosity.

\subsection{Pore-scale resolved model} \label{sec:porescale}

At the pore scale, fluid flow in the whole flow domain 
$\Omega_\text{flow}$ is governed by the Stokes equations, 
\begin{equation}\label{eq:pore}
\deld \vec v = 0, \qquad 
- \deld \ten T(\vec  v, p) =\vec 0 \qquad \text{in} \;\; \Omega_\text{flow},
\end{equation}
completed with the no-slip condition on the boundary of solid inclusions  
\begin{equation}\label{eq:no-slip}
\vec v = \vec 0 \quad \text{on} \;\; \partial \Omega_\text{flow} \setminus \partial \Omega,
\end{equation}
and appropriate conditions on the external boundary $\partial \Omega$. 
Here, $\vec v$ and $p$ denote the fluid velocity and pressure, $\ten T(\vec v,p) =\mu \nabla \vec v - p \ten I$ the stress tensor, $\ten I$ the identity tensor and $\mu$ the dynamic viscosity. 

Resolving pore-scale information is computationally expensive for practical applications. Therefore, REV-scale model formulations, which accurately reflect the pore-scale flow processes, are often preferred and are studied in this manuscript. The pore-scale resolved model~\eqref{eq:pore}--\eqref{eq:no-slip} is used only as a reference for the model validation purposes. 
A finite-volume scheme on staggered grids, also known as MAC scheme \cite{harlow1965a}, is used to discretize the pore-scale model~\eqref{eq:pore}--\eqref{eq:no-slip}.

\subsection{Subdomain models}\label{sec:subdomain}

In this study, we consider two different types of coupled models, for which the Stokes equations are used in the free-flow region $\Omega_\FF$ but the porous domain $\Omega_\PM$ is treated by different modeling concepts. The first type of model relies on the REV-scale description of the porous-medium domain using Darcy's law, whereas the second type of model follows a hybrid-dimensional approach, where a lower-dimensional pore-network model (PNM) is used to describe the fluid flow in the porous domain \cite{Weishaupt_2019, weishaupt2020a}. 

\subsubsection{Free-flow model}
As a common feature, both coupled models (REV-scale model, pore-network model) contain the incompressible, stationary Stokes equations for the description of fluid flow in the free-flow domain 
\begin{equation}\label{eq:Stokes}
\deld \vec{v}_\FF = 0, \quad 
- \deld \ten T(\vec v_\FF,p_\FF) = \vec 0
\qquad \text{in} \;\; \Omega_\FF,
\end{equation}
where $\vec v_\FF$ is the fluid velocity and $p_\FF$ is the fluid pressure. For discretization of the Stokes system~\eqref{eq:Stokes}, the same MAC scheme as for model \eqref{eq:pore}--\eqref{eq:no-slip} is employed.

\subsubsection{REV-scale porous-medium model}
At the REV-scale, fluid flow through the porous medium is described by the Darcy flow equations 
\begin{align}\label{eq:Darcy}
\deld \vec v_\PM = 0, \quad \ \
\vec v_\PM = -\frac{\ten K}{\mu} \del p_\PM \quad \text{in} \;\; \Omega_\PM, 
\end{align}
where $\vec v_\PM$ is the Darcy fluid velocity, $p_\PM$ 
is the fluid pressure and $\ten K$ is the intrinsic permeability tensor, which is symmetric, positive definite, and bounded.
Equations~\eqref{eq:Darcy} are discretized with a vertex-centered finite-volume scheme, also known as box method \cite{Hackbusch:1989}. This scheme has the advantage that degrees of freedom are naturally located at the interface and therefore directly allow the calculation of interfacial quantities (see~\cite{schneider2021coupling} for more details) appearing in the coupling conditions \eqref{eq:NEW-mass}--\eqref{eq:NEW-tangential} below.

\subsubsection{Pore-network porous-medium model}\label{sec:pnm}

Pore-network models \cite{blunt2017a} consider a simplified yet equivalent representation of the porous
geometry by separating the void space into larger pore bodies connected by narrow
pore throats. Despite their low computational demand, a rather high degree of pore-scale accuracy can be achieved \cite{oostrom2016a}.
PNMs can also be combined with modeling approaches on different scales \cite{scheibe2015a}, such as Darcy-type continuum models \cite{Balhoff2007b, Balhoff2007a, mehmani2014a} or free-flow models \cite{beyhaghi2016a}. Weishaupt et al.~\cite{Weishaupt_2019} developed a monolithic approach to couple a
pore-network model with a (Navier--)Stokes model, which was later improved by considering pore-scale slip~\cite{weishaupt2020a}.

\begin{figure}[!ht]
    \centering
    \includegraphics[scale=0.525]{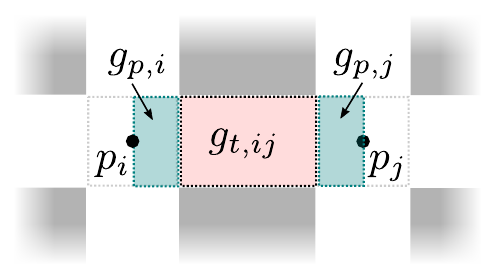} \hspace{1.5ex}
    \caption{Schematic contribution to total conduction for the PNM. Throat $ij$ connects the pore bodies $i$ and $j$ at the centers of which the pressures $p_i$ and $p_j$ are defined. $g_{t,ij}$ is the throat conductance valid for the region marked in light red. $g_{p,i}$ and $g_{p,j}$ are the conductances defined for the pore body halves marked in teal.}
    \label{fig:pnm}
\end{figure}

For the PNM, we require the conservation of mass for each pore body $i$ (the intersection of two or more pore throats):
\begin{equation}
\label{eq:pnm}
 \sum_j Q_{ij} = 0, \qquad Q_{ij} = g_{ij} (p_i - p_j).
\end{equation}
Here, $Q_{ij}$ is the discrete volume flow rate in a throat connecting the pore bodies $i$ and $j$, and the pressures defined at the centers of the pore bodies $i$ and $j$ are given by
$p_i$ and $p_j$ (Figure~\ref{fig:pnm}). Equation~\eqref{eq:pnm} represents  a  finite-volume discretization scheme with a two-point flux approximation, see~\cite{weishaupt2020a,koch2021} for further details.
The total conductance $g_{ij}$ is determined by the pore throat geometry and the fluid properties. 
Considering the pressure losses both within the pore bodies and throats, we use
\begin{equation}
\label{eq:pnmcond}
g_{i j}=\left(g_{t,i j}^{-1}+g_{p,i}^{-1}+g_{p,j}^{-1}\right)^{-1} ~,
\end{equation}
where $g_{t,i j}$ is the conductance of a throat $ij$ while $g_{p,i}$ and $g_{p,j}$ are the conductances
of the adjacent pore-body halves 
(Figure~\ref{fig:pnm}). Simple analytical expressions for $g_{ij}$ are available in the literature~\cite{patzek2001a} for certain geometries. Usually, we determine $g_{ij}$ via numerical upscaling \cite{mehmani2017a}, whereas for this study, we consider it to be an additional uncertain parameter. In the following, we only refer to $g_{p,i}$, as for the given geometry $g_{p,i} = g_{p,j}$ for interior throats. At interface throats, one of the half-pore-body conductance is zero.  

\subsection{Coupling concepts}
\label{sec:coupling-concepts}
A variety of REV-scale coupling concepts for the Stokes--Darcy system \eqref{eq:Stokes}--\eqref{eq:Darcy} is available in the literature. In this paper, we consider the most widely used set of interface conditions, based on the Beavers--Joseph condition, and the recently developed generalized conditions~\cite{Eggenweiler_Rybak_MMS20}. If the PNM \eqref{eq:pnm} is used in the porous medium, separate coupling conditions, suitable for the pore-scale description of interface exchange processes, must be considered.

\subsubsection{Classical coupling conditions (REV-scale model)}\label{sec:classical}

The most commonly used interface conditions are the \textit{conservation of mass}
\begin{equation}\label{eq:IC-mass}
\vec v_\FF \vdot \vec n = \vec v_\PM \vdot \vec n \qquad \text{on} 
\;\Gamma,
\end{equation}
the {\it balance of normal forces}
\vspace*{-0.5ex}
\begin{equation}\label{eq:IC-momentum}
-\vec n \vdot \ten T (\vec v_\FF, p_\FF)  \vec n = p_\PM 
\qquad \text{on} \; \Gamma,
\end{equation}
and the {\it Beavers--Joseph condition}~\cite{Beavers_Joseph_67} for the tangential component of velocity
\begin{equation}\label{eq:IC-BJJ}
(\vec v_\FF - \vec v_\PM ) \vdot \vec \tau
- \frac{\sqrt{\ten K}}{\alpha_\BJ}
\vec \tau \vdot \nabla \vec v_\FF \vec n = 0  \qquad \text{on} \; \Gamma.
\end{equation}
Here, $\alpha_\BJ>0$ is the Beavers--Joseph parameter, $\vec n$ is the normal unit vector on $\Gamma$ pointing outward from the porous medium,  $\vec \tau$ is a tangential unit vector on $\Gamma$ and $\sqrt{\ten K} = \sqrt{\vec \tau \vdot \ten K \vec \tau}$ (Figure~\ref{fig:setting}). 

The Beavers--Joseph interface condition~\eqref{eq:IC-BJJ} was postulated for flows parallel to the interface~\cite{Beavers_Joseph_67}. In~\cite{Eggenweiler_Rybak_JFM20, Eggenweiler_Rybak_MMS20} it is shown that this condition is unsuitable for arbitrary flow directions to the porous bed, however, it is routinely applied in the literature to multidimensional flows~\cite{Discacciati_Gerardo-Giorda_18, Hanspal_etal_09}. 
The Stokes--Darcy problem~\eqref{eq:Stokes}--\eqref{eq:Darcy},~\eqref{eq:IC-mass}--\eqref{eq:IC-BJJ} contains uncertain model parameters. 

In this paper, we focus on four uncertain parameters: the exact interface position $\gamma$, the Beavers--Joseph slip coefficient $\alpha_\BJ$, the permeability tensor $\ten K = k \ten I$ and the maximum boundary velocity at the inflow boundary $V^{\text{top}}$ (Fig.~\ref{fig:BC}). The exact location of the fluid--porous interface for REV-scale models is not known a priori. In the literature, there are several recommendations to impose the sharp interface directly on the top of solid inclusions in case of circular grains~\cite{Lacis_etal_20, Rybak_etal_20,Beavers_Joseph_67}. For other pore geometries, especially in the case of anisotropic  media, the problem of optimal interface location is still open. However, the exact interface location is very important, especially  for microfluidic models~\cite{Terzis_etal_19} which are routinely used in experiments. 
Another uncertain parameter is the Beavers--Joseph coefficient $\alpha_\BJ$,  which is supposed to contain the information on the surface roughness~\cite{Beavers_Joseph_67, Bars_Worster_06}. An investigation to calibrate this parameter was recently carried out in~\cite{Rybak_etal_20}, however, only for isotropic media. There was also an attempt to determine the Beavers--Joseph coefficient experimentally for flows parallel to the fluid--porous interface, isotropic and orthotropic porous media~\cite{Mierzwiczak_19}, where the Beavers--Joseph parameter was found to be $\alpha_\BJ < 1$ and dependent on the intrinsic permeability. 
Finally, the permeability tensor appearing in the Beavers--Joseph condition \eqref{eq:IC-BJJ} is not necessarily the permeability of the porous bulk, as in the standard models~\cite{Discacciati_Miglio_Quarteroni_02, Discacciati_Quarteroni_09}, but could also be permeability of the near-interfacial region~\cite{Lacis_Bagheri_17, Zampogna_Bottaro_16}.

\subsubsection{Generalized coupling conditions for arbitrary flows (REV-scale model)}
\label{sec:new}

An alternative to the classical interface conditions for Stokes--Darcy problems are the
generalized coupling conditions derived rigorously in~\cite{Eggenweiler_Rybak_MMS20} via homogenization and boundary layer theory.
These conditions are valid for arbitrary flow directions to the fluid--porous interface and read
\begin{alignat}{2}
    \vec v_\FF \vdot \vec n &= \vec v_\PM \vdot \vec n \quad \ &&\text{on } \Gamma, \label{eq:NEW-mass}
    \\[1ex]
    p_\PM &= -\vec n \vdot \ten T (\vec v_\FF, p_\FF) 
    \vec n \notag
    \\ & \quad \
    + \mu N_s^\text{bl}  \vec \tau \vdot \nabla \vec v_\FF
    \vec n \quad \ &&\text{on } \Gamma, \label{eq:NEW-momentum}
    \\
    \vec v_\FF \vdot \vec \tau &= 
    - \ell N_1^\text{bl}  \vec  \tau \vdot
    \nabla \vec v_\FF
    \vec n \notag 
    \\ & \quad \ + 
    \mu^{-1} \ell^2 \sum_{j=1}^2 \frac{\partial p_\PM}{\partial x_j} \vec M^{j,\text{bl}}
    \vdot \vec \tau  \quad \ &&\text{on } \Gamma. \label{eq:NEW-tangential}
\end{alignat}
Here, $\vec M^{j,\text{bl}}$, $N_1^\text{bl}$ and $N_s^\text{bl}$ are boundary layer constants introduced in~\cite{Eggenweiler_Rybak_MMS20}.
For the generalized coupling conditions the interface can be located at the distance  
$\mathcal{O}(\ell)$ from the top of the first row of solid inclusions, where $\ell$ denotes the characteristic pore size. Based on the chosen interface position and the pore geometry, the effective coefficients appearing in conditions~\eqref{eq:NEW-mass}--\eqref{eq:NEW-tangential} are computed numerically using the theory of homogenization and boundary layers~\cite{Carraro_etal_15, Hornung_97, Jaeger_Mikelic_00, Jaeger_Mikelic_09}.
This is the main advantage of the generalized interface conditions, besides their suitability for arbitrary flows in coupled Stokes--Darcy systems.
There are also other coupling concepts for Stokes--Darcy problems in the literature, e.g.,~\cite{Angot_etal_17, AGOT2020, Lacis_etal_20,OchoaTapia_Whitaker_95}, which are beyond the scope of this study.

\subsubsection{Coupling conditions for the pore-network model}\label{sec:pn}

\begin{figure}[h]
    \centering
    \includegraphics[scale=0.26]{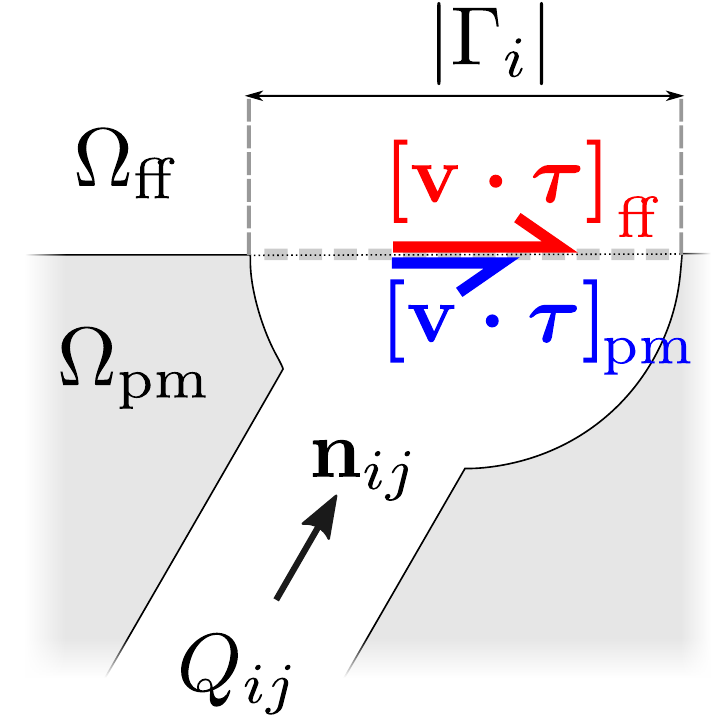}
    \caption{Schematic representation of local interface for the free-flow/PNM. 
    }
    \label{fig:slip_vel_pnm}
\end{figure}

Each intersection of a pore body $i$ with the free-flow domain boundary yields a pore-local discrete interface $\Gamma_i$ on which we formulate coupling conditions (Figure~\ref{fig:slip_vel_pnm}). We assume no-flow/no-slip condition for the free flow at the location of solid grains (no intersecting pore throat). This results in the following coupling conditions for the free-flow/pore-network model
\begin{align}
    \vec v_\FF \vdot \vec n &= \vec v_\PM \vdot \vec n \qquad \text{on } \Gamma_i~, \label{eq:pnm-mass}
    \\[1ex]
    p_\PM &=  p_\FF \hspace*{1.3cm} \text{on } \Gamma_i ~, \label{eq:pnm-momentum}
    \\
     \vec v_\FF \vdot \vec \tau
     &= \begin{cases}
v_{\mathrm{slip}} \hspace*{0.8cm}\text{on } \Gamma_i ~,\\
0 \hspace*{1.25cm}\text{else} ~, \label{eq:pnm-slip}
\end{cases}
\end{align}
with 
\begin{equation}
\label{eq:v_slip}
 v_{\mathrm{slip}} \!=\! -\frac{1}{\beta_\pore} \left[ (\nabla{ \vec v} \!+ \! \nabla{ \mathbf{v}}^T)  \vec n \vdot \vec \tau \right]_\FF \!+ \! \left[ \mathbf{v} \vdot  \vec \tau \right]_\PM.
\end{equation}

We approximate the tangential component of the pore-body interface velocity as
\begin{equation}
\label{eq:v_throat_x}
[\vec v \vdot \vec \tau]_\PM = 
\frac{Q_{ij}}{\lvert \Gamma_{i} \rvert} 
[ \mathbf{n}_{ij} \vdot \vec{\tau}]_\PM~,
\end{equation}
where $\mathbf{n}_{ij}$ is a unit normal vector parallel to the throat's central axis and pointing towards the interface $\Gamma_i$. The volume flow through the pore throat $ij$ is given by $Q_{ij}$ and $\lvert\Gamma_i\rvert$ is the area of the discrete coupling interface.

Equations~\eqref{eq:pnm-slip} and~\eqref{eq:v_slip} can be seen as the pore-scale analog to Equation~\eqref{eq:IC-BJJ} with a pore-local slip coefficient $\beta_\pore$,
which is determined numerically in a preprocessing step. We refer to \cite{weishaupt2020a} for more details.
The three sets of coupling conditions~\eqref{eq:IC-mass}--\eqref{eq:IC-BJJ},~\eqref{eq:NEW-mass}--\eqref{eq:NEW-tangential} and~\eqref{eq:pnm-mass}--\eqref{eq:pnm-slip} are discretized corresponding to the adjacent subdomain models' discretizations, and the resulting coupled discrete models are treated by a monolithic strategy, assembling all contributions in a single system of equations for each model.

\section{Benchmark scenario}\label{sec:scenarios}

Corresponding to Figures~\ref{fig:setting} and~\ref{fig:BC}, we investigate rectangular solid inclusions of size $d$ and study a flow problem where the flow is arbitrary to the fluid--porous interface $\Gamma$. We consider laminar fluid flow through the coupled flow domain with viscosity $\mu = 10^{-3}$ Pa $\cdot$ s. We describe the geometrical configuration and the boundary conditions in~Section~\ref{sec:square-BC}, followed by a description of the uncertainties in~Section~\ref{sec:square-UQ} and the system response quantities in~Section~\ref{sec:square-response}.

\subsection{Geometrical setting and boundary conditions} \label{sec:square-BC}
\begin{figure}[htb]
    \centering
    \quad \includegraphics[scale=0.7]{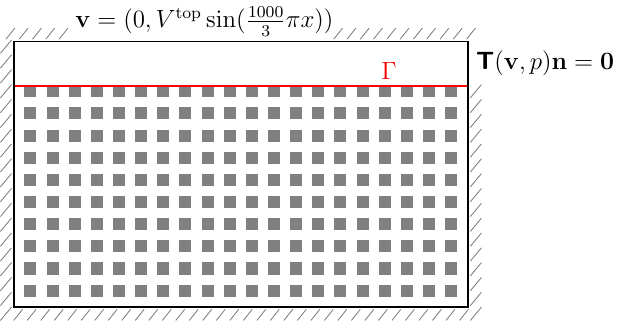} \\[2ex]  \includegraphics[scale=0.45]{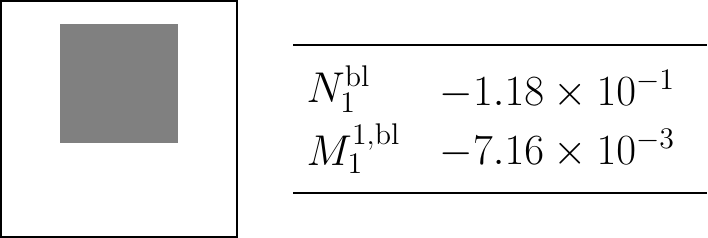} \\[1ex]
    \caption{Schematic description of the coupled flow problem (top), 
    unit cell and non-dimensional effective parameters for the interface location $\gamma=5.05\,\mm$ (bottom). 
    }
    \label{fig:BC}
\end{figure}

We consider the free-flow region $\Omega_\FF = (0, L) \times (\gamma, H)$ and the porous-medium domain $\Omega_\PM = (0, L) \times (0, \gamma)$ with $L = 10.25\,\mm$ and $H = 6\,\mm$, separated by the sharp fluid--porous interface $\Gamma = (0,L)\times \{\gamma\}$, where the value for $\gamma$ is uncertain. 
The porous medium is isotropic, $\ten K = k \ten I$, and consists of $20 \times 10$ square solid inclusions of size $d = 0.25\,\mm$ (Figure~\ref{fig:BC}) leading to porosity $\phi=0.75$. The inclusions are positioned in such a way that the line tangent to the top of the upper row of solid inclusions is given by $(0, L) \times\{5\,\mm\}$ and the characteristic pore size appearing in coupling condition~\eqref{eq:NEW-tangential} is $\ell= 0.5\,\mm$. 

For the classical interface conditions \eqref{eq:IC-mass}--\eqref{eq:IC-BJJ} (\textit{Classical IC}) the Beavers--Joseph parameter is typically taken $\alpha_\BJ=1$ in the literature, although it is often not the optimal choice~\cite{Rybak_etal_20,Mierzwiczak_19,Lacis_etal_20}. Here, we consider $\alpha_\BJ$ as an uncertain parameter, which is quantified in~Sections~\ref{sec:square-UQ} and~\ref{sec:square-response}.
The boundary layer constants appearing in the generalized coupling conditions~\eqref{eq:NEW-mass}--\eqref{eq:NEW-tangential} (\textit{Generalized IC}) are computed numerically based on the geometrical configuration of the interfacial zone and are presented in~Figure~\ref{fig:BC} (bottom). For details on the computation of these effective parameters, we refer the reader to~\cite{Eggenweiler_Rybak_MMS20}.
Note that the boundary layer constants $N_1^\mathrm{bl}$ and $M_1^\mathrm{bl}$ (Figure~\ref{fig:BC}, bottom) are non-dimensional. For isotropic porous media, the constants $M_1^\text{2,bl}=0$ and $N_s^\text{bl}=0$, therefore, they do not appear in~Figure~\ref{fig:BC} (bottom). 

In order to obtain a closed formulation for the pore-scale problem~\eqref{eq:pore}--\eqref{eq:no-slip} we set the following boundary conditions on the external boundary
\begin{align}\label{eq:pore-scale-BC}
    &\vec v = \vec v_\text{in} =(0, V^\text{top} \operatorname{sin}(\tfrac{1000}{3}\pi x)) && \text{ on } \Gamma_\text{in},
    \\
    &\ten T(\vec v,p)\, \vec n_\FF = \vec 0 && \text{ on } \Gamma_\text{out}, 
    \\
    &\vec v = \vec 0  && \text{ on } \Gamma_\text{nf}, 
\end{align}
where the inflow boundary $\Gamma_\text{in} = (3\,\mm, 6\,\mm) \times \{H\}$, $\Gamma_\text{out}= \{L\} \times (5.5\, \mm, H) $, $\Gamma_\text{nf} = \partial \Omega \setminus (\Gamma_\text{in} \cup \Gamma_\text{out})$ and $\vec n_\FF$ is the outward unit normal vector on $\partial \Omega_\FF$.
The boundary conditions for the coupled flow problem are presented schematically in Figure~\ref{fig:BC} (top).

The corresponding boundary conditions for the REV-scale model formulation given by \eqref{eq:Stokes}--\eqref{eq:Darcy} together with either the \textit{Classical IC}~\eqref{eq:IC-mass}--\eqref{eq:IC-BJJ} or the \textit{Generalized IC}~\eqref{eq:NEW-mass}--\eqref{eq:NEW-tangential} read
\begin{alignat}{2}
    &\vec v_\FF = (0, V^\text{top} \operatorname{sin}(\tfrac{1000}{3}\pi x)) \qquad && \text{ on } \Gamma_\text{in}, \label{eq:macro-BC-1} 
    \\
    &\ten T(\vec v_\FF,p_\FF)\, \vec n_\FF = \vec 0  && \text{ on } \Gamma_\text{out}, \label{eq:macro-BC-2} 
    \\
    &\vec v_\FF = \vec 0 && \text{ on } \Gamma_\text{nf,ff}, \label{eq:macro-BC-3} 
    \\
    &\vec v_\PM \vdot \vec n_\PM = 0 && \text{ on } \Gamma_\text{nf,pm}, \label{eq:macro-BC-4}
\end{alignat}
where $\Gamma_\text{nf,ff} = \partial \Omega_\FF \setminus (\Gamma_\text{in} \cup \Gamma_\text{out} \cup \Gamma)$, $\Gamma_\text{nf,pm} = \partial \Omega_\PM \setminus  \Gamma$ and $\vec n_\PM$ denotes the unit normal vector on $\partial \Omega_\PM$ pointing outward the porous medium.

The boundary conditions~\eqref{eq:macro-BC-1}--\eqref{eq:macro-BC-3} also hold for the hybrid-dimensional free-flow/pore-network model (\textit{Pore-Network model}), such that no mass enters or leaves the domain through the pores on $\Gamma_\text{nf,pm}$. The coupling conditions~\eqref{eq:pnm-mass}--\eqref{eq:pnm-slip} are set on $\Gamma$ for PNM.

\subsection{Uncertainties and errors}\label{sec:square-UQ}
So far, we have presented various coupling concepts for free flow and porous-medium flow. These concepts differ in the description of processes in the porous region and across the fluid--porous interface, where different mathematical models are employed, whereas the free flow for all concepts is modeled based on the stationary Stokes equations. The uncertainty, due to the choice of adequate representation of the system of interest, is known as conceptual uncertainty. 
In addition to the conceptual uncertainty, each computational model includes uncertain parameters, such as material parameters, or interface location, requiring a thorough investigation. This type of uncertainty is known as parametric uncertainty. Uncertain model inputs, defined later, must be propagated through the model or simulation (also known as uncertainty propagation) to effectively assess competing modeling concepts' response quantities and validate the corresponding computational models against a reference solution.

As described in Section~\ref{sec:classical}, regarding the coupled Stokes--Darcy model with the \textit{Classical IC}, the benchmark scenario contains four uncertain parameters: the maximum boundary velocity at the inflow boundary, the exact interface position, the Beavers--Joseph coefficient, and the permeability tensor.
In contrast to the \textit{Classical IC}, the \textit{Generalized IC} do not contain the Beavers--Joseph coefficient. Further, the \textit{Generalized IC} rely on the assumption that the interface location may not be below the top of the solid inclusions.
Correspondingly, the parameters and their associated distributions as prior knowledge for the Stokes--Darcy model with the \textit{Classical IC} and the \textit{Generalized IC} are listed in Tables~\ref{tab:Params_Ranges_StokesDarcy_classic} and \ref{tab:Params_Ranges_StokesDarcy_general}, respectively.
\begin{table*}[!ht]
\caption{The uncertain parameters and their defined distributions for the classical coupled Stokes--Darcy model.}
\centering
\begin{tabular}{l c c c}
\hline
 Parameter name & Range & Unit & Distribution type  \\
\hline
  Boundary velocity, $V^\text{top}$ & $\left[5\times 10^{-4}, 1.5\times10^{-3}\right]$ & $\m/\text{s}$ & uniform \\
  Exact interface location, $\gamma$ &$\left[4.9, 5.1\right]$ & $\mm$ & uniform \\
  Permeability, $k$  & $\left[10^{-9}, 10^{-9}\right]$ & $\m^2$ & uniform \\
  Beavers--Joseph parameter, $\alpha_\BJ$ & $\left[0.1,4\right]$ & - & uniform \\ 
\hline
\end{tabular}
\label{tab:Params_Ranges_StokesDarcy_classic}
\end{table*}

\begin{table*}[!ht]
\caption{The uncertain parameters and their associated distributions for the Stokes--Darcy model with the generalized interface conditions.}
\centering
\begin{tabular}{l c c c}
\hline
 Parameter name & Range & Unit & Distribution type  \\
\hline
  Boundary velocity, $V^\text{top}$ & $\left[5\times 10^{-4}, 1.5\times10^{-3}\right]$ & $\m/\text{s}$ & uniform \\
  Exact interface location, $\gamma$ &$\left[5.0, 5.1 \right]$ & $\mm$ & uniform \\
  Permeability, $k$  & $\left[10^{-10}, 10^{-8}\right]$ & $\m^2$ & uniform \\
\hline
\end{tabular}
\label{tab:Params_Ranges_StokesDarcy_general}
\end{table*}

As for the \textit{Pore-Network} model, we consider the total conductance $g_{ij}$ in \eqref{eq:pnmcond} (see Figure~\ref{fig:pnm}) as uncertain parameter to be inferred during the calibration phase. This parameter plays the role of permeability in the pore-network setting.
Another uncertain input parameter is the pore-scale slip coefficient $\beta_\pore$. It can be determined numerically in a preprocessing step, in which it is approximated by solving a simplified, equivalent problem of free flow over a single pore throat intersecting with the lower boundary of the free-flow channel~\cite{weishaupt2020a}. 
The list of considered uncertain parameters and their associated distribution as prior knowledge for the PNM are presented in Table~\ref{tab:Params_Ranges_PNM}.

\begin{table*}[!ht]
\caption{The uncertain parameters and their specifications for the pore-network model.}
\centering
\begin{tabular}{l c c c}
\hline
 Parameter name & Range & Unit & Distribution type  \\
\hline
  Boundary velocity, $V^\text{top}$ & $\left[5\times 10^{-4}, 1.5\times10^{-3}\right]$ & $\m/\text{s}$ & uniform \\
  Total conductance, $g_{ij}$ & $\left[10^{-7}, 10^{-5}\right]$ & $\m^3/\left(\text{s} \cdot \text{Pa}\right)$ & uniform \\ 
  Pore-local slip coefficient, $\beta_\pore$ & $[10^{3}, 10^{5}]$ & $1/\m$ & uniform \\
\hline
\end{tabular}
\label{tab:Params_Ranges_PNM}
\end{table*}

As opposed to uncertainties, errors are defined as the difference between the true value and the predicted value, and have both a sign and a magnitude. 
We consider the errors associated with the model discrepancy error, numerical approximation, and surrogate modeling in our analysis. These errors are aggregated and used as diagonal entries of the residual covariance matrix $\Sigma$ in the likelihood function in \eqref{eq:BayesLikelihood}.

\subsubsection{Model discrepancy error} 
In the current benchmark case study, we have a fully resolved pore-scale solution as a reference.  However, in practice, the reference solution is not available, and instead, experimental data is incorporated that includes an observation error. Nevertheless, the analyzed models in our benchmark study could never perfectly reproduce the ground truth, i.e., the reference solution. This difference can be attributed to the presence of model discrepancy. This discrepancy always exists for various reasons, such as simplified assumptions, missing physics, upscaling due to scale differences. 
Several methods have been proposed in the literature to incorporate the model discrepancy in a Bayesian setting. These methods' treatment of model discrepancy range from a constant bias to more sophisticated methods in which one forms a Bayesian hierarchical model to solve a joint parameter and model discrepancy inference problem \cite{kennedy2001bayesian,bayarri2007framework,brynjarsdottir2014learning,ling2014selection,gardner2021learning}.
Since the model discrepancy is not perfectly known, we parameterize it as $\Sigma(\theta_{\epsilon})$ and treat its parameter $\theta_{\epsilon}$ as additional unknown parameter. Following~\cite{wagner2021uqlabbayes}, we infer these parameters jointly with model parameters $\theta_k$ in \eqref{eq:BayesLikelihood}. 
We consider a diagonal covariance matrix as $\Sigma = \sigma^2 \mathbb{I}_{N_\text{out}}$ with a scalar unknown parameter $\sigma^2$ for each system response quantity, i.e., velocity and pressure (Section~\ref{sec:square-response}).
$N_\text{out}$ stands for the number of data points.

\subsubsection{Numerical error}
The governing equations of the models under investigation in this study require approximation of numerical solutions. These approximations provide an additional source of error.
There are five primary sources of errors in computational physics solutions, given that the numerical scheme is stable, consistent, and robust. These sources are insufficient spatial discretization, insufficient temporal discretization for unsteady-state flow, insufficient iterative convergence, computer round-off, and computer programming \cite{roy2019errors}.
Since quantifying errors from these sources is the main focus in the verification of numerical schemes, we only investigate the discretization error that originates from a certain choice of mesh size. Following~\cite{oberkampf2010verification}, we take a heuristic approach to quantify this error, in that we fit generalized Richardson extrapolation to estimate the error by comparing three different mesh spacings. The Richardson extrapolation takes the following form
\begin{equation}
\label{eq:RichardsonExtrapol}
f_{k}=\bar{f}+e_{p} h_{k}^{\hat{p}} + \mathcal{O}\left(h_{k}^{\hat{p}+1}\right),
\end{equation}
%
\noindent
where $f_{k}$ denotes the exact solution to the discrete equation on a mesh with a known spacing $h_{k}$, $\bar{f}$ stands for the exact solution to the original PDE (unknown), $e_{p}$ is the error term coefficient, and $\hat{p}$ indicates the observed order of accuracy. Here, we seek the first-order error. Thus, the unknowns $\bar{f}$ and $e_{p}$ can be easily determined via the least square method for the numerical solutions obtained by varying the mesh spacing.

\subsubsection{Surrogate error}
As previously mentioned, we substitute the computational models with the easy-to-evaluate surrogate models in the Bayesian analysis to offset the computational cost. This replacement also introduces a new source of error, known as a surrogate prediction error. Ignoring this error could result in a biased posterior distribution. As for prediction uncertainty, a mean squared error based on a testing set can provide a good estimate of the surrogate error variance.

We incorporate the errors discussed above in the Markov Chain Monte Carlo (MCMC) simulation method to approximate the posterior distribution \cite{robert2013monte,liu2008monte}, used in the calibration stage. We directly sum up all the covariance matrices of errors to obtain the likelihood calculations' total covariance matrix. 
Here, we assume that all these errors follow a normal distribution and are independent of each other.

\subsection{System response quantities}\label{sec:square-response}
\begin{figure}[htb]
    \includegraphics[scale=0.65]{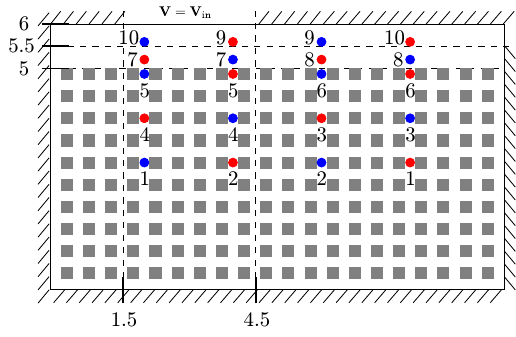} 
    \quad
    \includegraphics[scale=0.65]{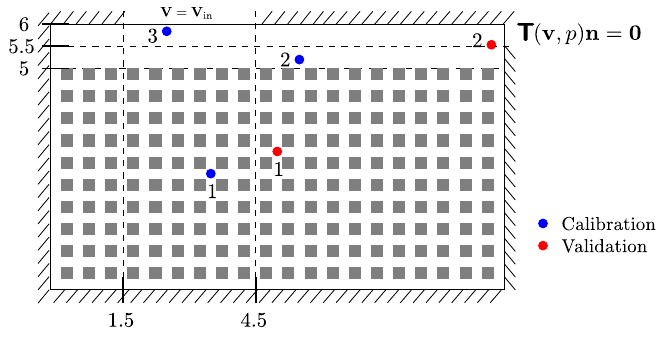} 
    \caption{Data extraction points for the velocity field (top) and pressure 
    (bottom) for the 
    calibration and validation scenarios.}
    \label{fig:data-extraction-points}
\end{figure}

A fundamental ingredient of each benchmark is the definition of so-called system response quantities (SRQs). These quantities define the prescribed output from the reference/experimental data as well as from the computational models that are compared in terms of the validation metric. The SRQs can be either local or global quantities. While the former can take quantities within the solution domain on the PDEs, such as dependent variables of the PDEs, the latter represents integral quantities or net flux out of a system.
As part of model validation, we seek to compare system responses generated by  different coupling concepts with the ones from the pore-scale resolved model~(Section~\ref{sec:porescale}).
Figure~\ref{fig:data-extraction-points} shows the data extraction points for the velocity field (top) and the pressure (bottom). 
Since less variability is expected for pressure values, we have selected fewer extraction points for pressure responses.
We train the surrogate models for all computational models based on the simulation results for the marked points. The points colored in blue and red provide the corresponding data for the calibration and validation steps, respectively.

The pore-scale resolved simulation results contain both macroscopic and microscopic details of the flow field. The latter become visible as oscillations of the pore-scale solutions in the porous medium. 
To make numerical simulation results comparable we need to average them at the pore and REV scale.
We consider volume averaging, where the averaged velocity field at a given point $\vec x_0 \in \Omega$ is obtained as
\begin{align}
    \vec v_\text{avg} (\vec x_0) = \frac{1}{\vert V(\vec x_0)\vert}\int_{V_\text{f}(\vec x_0)} \vec v (\vec x)\, \text{d} \vec x \, ,
\end{align}
where $V(\vec x_0)$ is the representative elementary volume corresponding to $\vec x_0$ and $V_\text{f}(\vec x_0)$ is its fluid part. The representative elementary volume $V(\vec x_0)$ has the same size as the periodicity cell $ Y^\ell$. Moreover, also the simulation results in the free-flow region need to be averaged correspondingly such that the interpretation of the SRQs is the same in all cases.

\section{Application of the Bayesian framework}
\label{sec:results}
In this section, we compare the coupled models (using either REV-scale formulation or \textit{Pore-Network} model in $\Omega_\PM$) with the pore-scale resolved model~\eqref{eq:pore}--\eqref{eq:no-slip}. 
For the REV-scale model formulation, we consider the Stokes--Darcy problem with the \textit{Classical IC} \eqref{eq:IC-mass}--\eqref{eq:IC-BJJ} and the \textit{Generalized IC}~\eqref{eq:NEW-mass}--\eqref{eq:NEW-tangential}.
We demonstrate that the latter ones are more accurate than the \textit{Classical IC} in case of parallel flows to the interface and that they are suitable for arbitrary flow directions where the \textit{Classical IC} fail.
As shown in~\cite{Weishaupt_2019, weishaupt2020a}, the hybrid-dimensional coupled model using a \textit{Pore-Network} approach in $\Omega_\PM$ can be an efficient and accurate choice  for simulating free flow over structured porous media.

\begin{figure}[t]
    \qquad \includegraphics[scale=0.15]{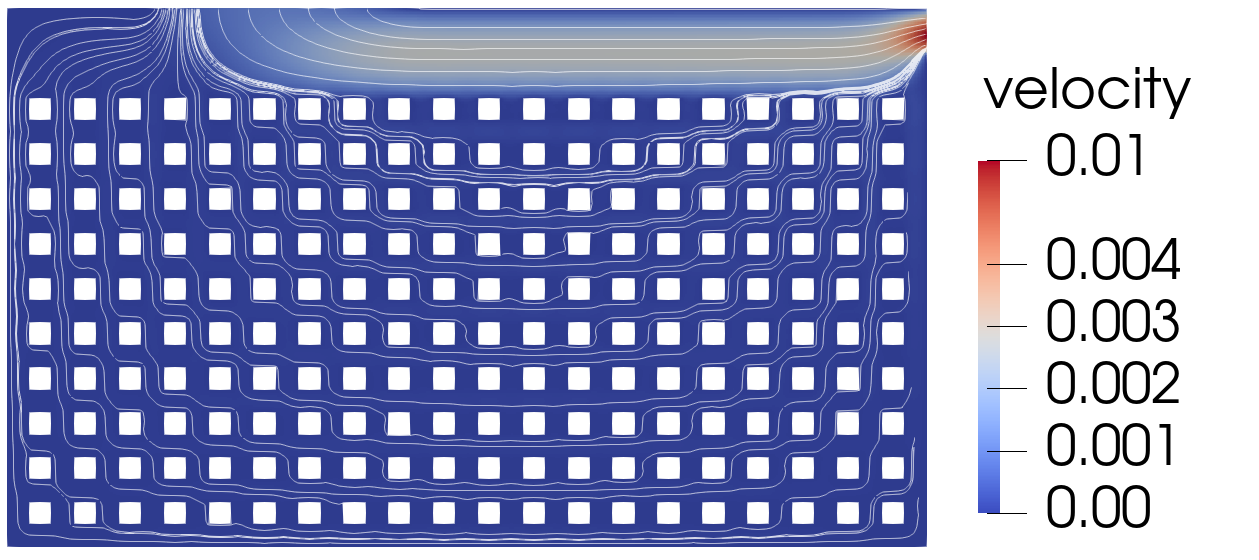} 
    \\[1ex]
    \hspace*{0.01cm}\qquad \includegraphics[scale=0.15]{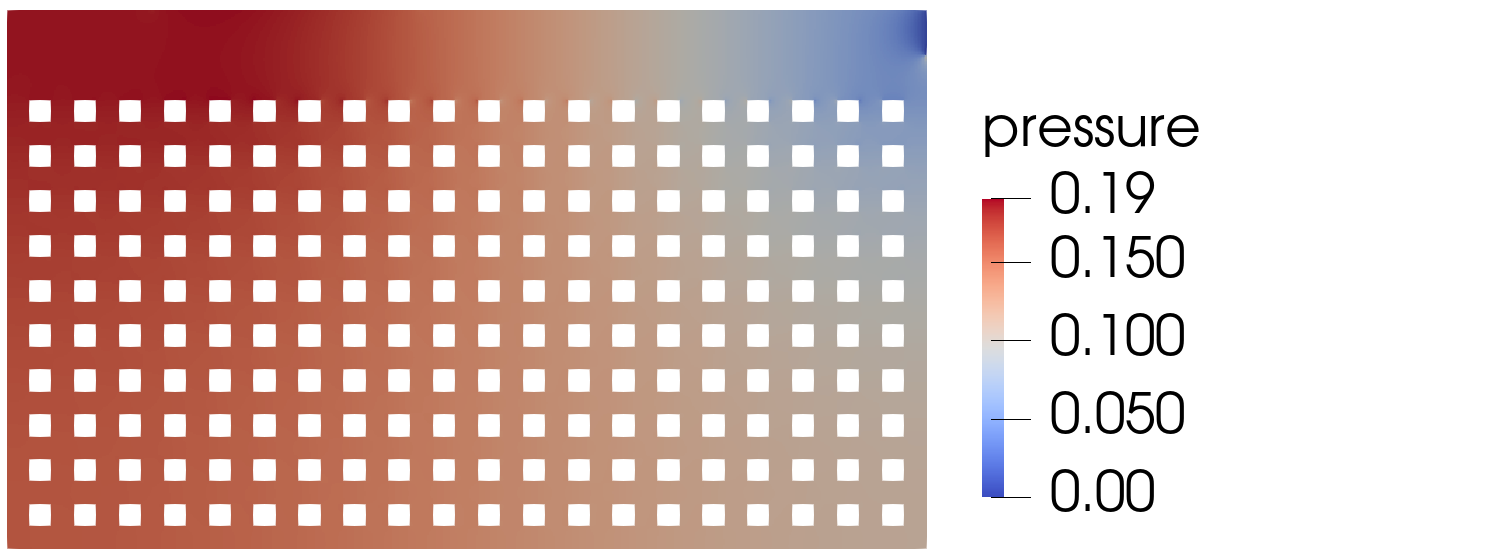} 
    \caption{Streamlines of the pore-scale velocity field (top) and the pressure (bottom).}
    \label{fig:streamlines}
\end{figure}

Here, our goal is to assess the coupled model's accuracy compared to the above-mentioned REV-scale approaches and under the influence of pore-scale parameter uncertainty. As reference data, we use the fully resolved pore-scale model for the velocity and pressure (Figure~\ref{fig:streamlines}). However, it is worth mentioning that the Stokes--Darcy model with \textit{Classical IC} and \textit{Generalized IC} can only offer predictions on the REV scale. Therefore, we average the values of SRQs obtained for the fully resolved pore-scale model as well as the \textit{Pore-Network} model for consistency. The averaging is performed via a volume averaging approach, discussed in Section~\ref{sec:square-response} to make the REV-scale numerical simulation results comparable with that of the pore-scale resolved simulation. 

The pore-scale model and the coupled models have been implemented in the open-source simulator DuMu$^\textrm{x}$ \cite{koch2021dumux}. 
The Bayesian analysis has been performed using a new open-source, object-oriented Python package \textit{BayesValidRox}\footnote{\url{https://pypi.org/project/bayesvalidrox/}}. It provides an automated workflow for surrogate-based sensitivity analysis, Bayesian calibration, and validation of computational models with a modular structure.

Replacing the models with their surrogates drastically reduces the total computational time of the analysis. This gain is essential in computationally demanding uncertainty quantification tasks, such as propagation or inference. In this study, we observed that by using a well-trained surrogate model, we could speed up one simulation run from $10 \sim 15$ s to only $0.005 \sim 0.007$ s with acceptable accuracy.  

The current section offers insights into analysis of predictive abilities in Section~\ref{sec:pred_abilities} and model comparison in Section~\ref{sec:model_comparison} using the surrogate-based Bayesian validation framework introduced in Section \ref{sec:bayes}. Additionally, we assess the influence of various modeling parameters onto the final model prediction, performing the global sensitivity analysis in Section~\ref{sec:sensitivity}.

\subsection{Global sensitivity analysis}
\label{sec:sensitivity}
In this section, we analyze how the variability of the model response quantities introduced in Section~\ref{sec:square-response} at the selected data extraction points (Figure~\ref{fig:data-extraction-points}) is affected by the variability of each input variable or combinations thereof. This is achieved via a sensitivity analysis. Various sensitivity analysis approaches have been developed in recent years. For an extensive review of different techniques, we refer the reader to~\cite{iooss2015review}. In the current paper, we explore the connection of polynomial representation to a global sensitivity measures \cite{oladyshkin2012global} and use the so-called \textit{Sobol indices} \cite{sobol1993sensitivity}, derived from a variance decomposition of model outputs in terms of contributions of each input parameter or combinations thereof. 
Using the Sobol decomposition, one can describe the total variance of the model in terms of the sum of the summands' variances. This variance decomposition is extensively explained in~\cite{sudret2008global}. Leveraging the orthonormality of the polynomial chaos basis, the authors also derive the so-called PC-based Sobol indices. The idea behind these indices is as follows: once the PC representation of the model in \eqref{eq:PCE_Trunc} is available, the expansion coefficients $c_{\vec{\alpha}}$ are simply gathered according to the dependency of each basis polynomial, square-summed and normalized
\begin{equation}
\label{eq:pce-sobol-1st}
\begin{array}{l}
S_{i_{1}, \ldots, i_{s}}=\frac{\sum\limits_{j=1}^{M} \chi_{j} c_{j}^{2}}{\sum\limits_{j=1}^{M} c_{j}^{2}} \, ,\\[2mm]
\chi_{j}=\left\{\begin{array}{ll}
1, & \text { if } \alpha_{j}^{k}>0, \; \forall j \in\left(i_{1}, \ldots, i_{s}\right) \\[0.5em]
0, & \text { if } \alpha_{j}^{k}=0, \; \exists j \in\left(i_{1}, \ldots, i_{s}\right)
\end{array}\right\} \, .
\end{array}
\end{equation}
%
Here,  $S_{i_{1}, \ldots, i_{s}}$ is the Sobol index that indicates what fraction of total variance of the response quantity can be traced back to the joint contributions of the parameters $\theta_{i_{1}}, \ldots, \theta_{i_{s}}.$ The index selection operator $\chi_{j}$ indicates where the chosen parameters $\theta$ numbered as $i_{1}, \ldots, i_{s}$ (i.e., $\left.\theta_{i_{1}}, \ldots, \theta_{i_{s}}\right)$ have concurrent contributions to the variance within the overall expansion. Simply put, it selects all polynomial terms with the specified combination $i_{1}, \ldots, i_{s}$ of model parameters.

A complementing measure for sensitivity analysis is the \textit{Sobol Total Index}. It expresses the total contribution to the variance of model output due to the uncertainty of an individual parameter $\theta_j$ in all cross-combinations with other parameters
\begin{equation}
\label{eq:pce-sobol-total}
S_{j}^{T}=\sum_{\left\{i_{1}, \ldots, i_{s}\right\} \supset j} S_{i_{1}, \ldots, i_{s}},
\end{equation}
where $S_{j}^{T}$ is a summation of all Sobol indices in which the variable $\theta_j$ appears as univariate as well as joint influences.
The total Sobol index can take values larger than 1
when the impact of interactions among parameters on the total output variance is not negligible.
This characteristic is particularly prominent in highly
nonlinear problems like the ones investigated in this study.
In what fallows, we present the total Sobol indices for the SQRs and the data extraction points defined in Section~\ref{sec:scenarios}, for all three models for the calibration scenario.

\subsubsection{Classical IC}\label{sec:sens_cic} 
\begin{figure*}[htb!]
    \centering
    \includegraphics[width=.45\linewidth]{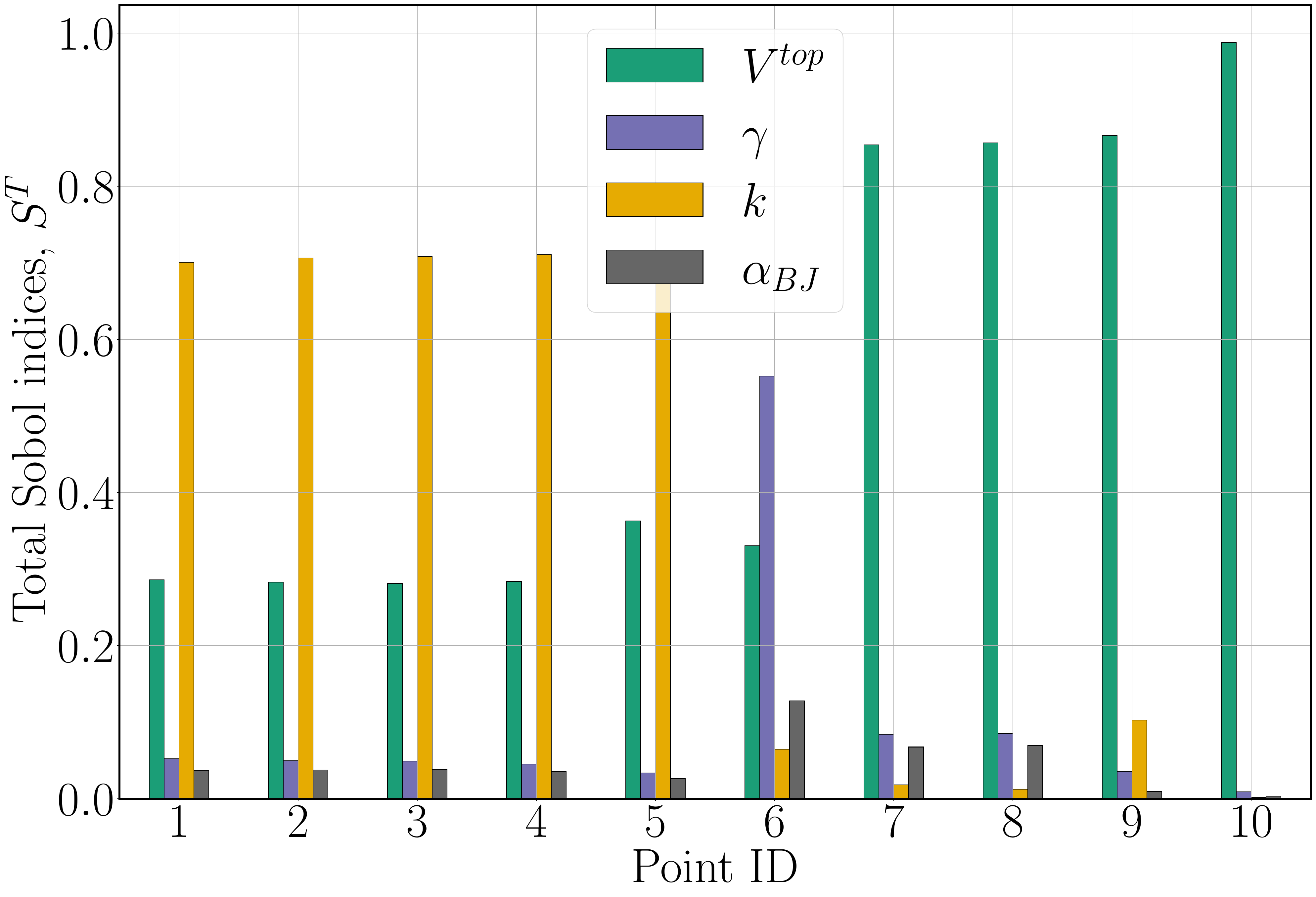} 
    \
    \includegraphics[width=.45\linewidth]{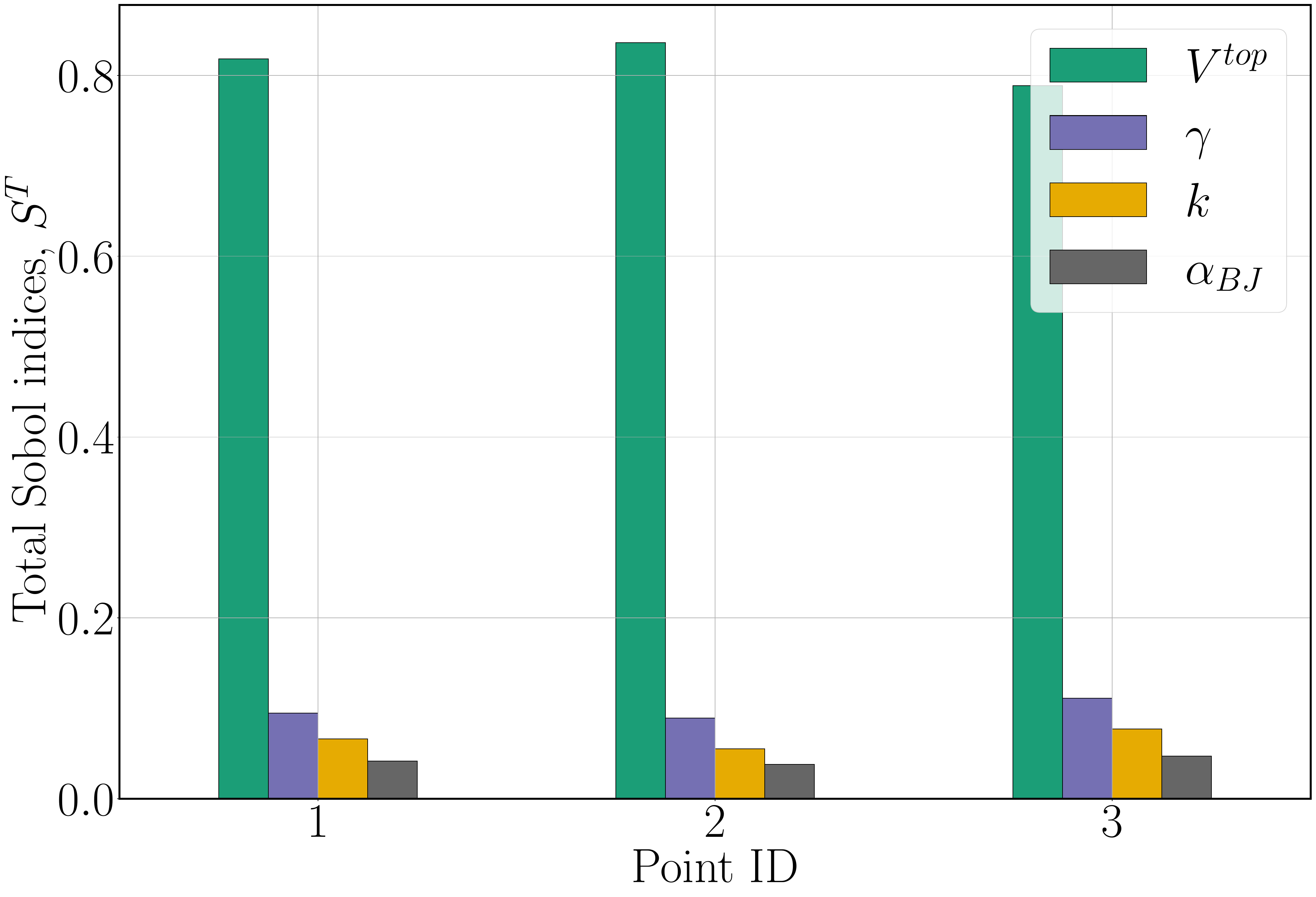} 
    \caption{Total Sobol indices of the Stokes--Darcy model with the \textit{Classical IC} for the velocity (left) and pressure (right) for the calibration (blue) points in Figure~\ref{fig:data-extraction-points}.}
    \label{fig:sensitivity_classical}
\end{figure*}

In~Figure~\ref{fig:sensitivity_classical}, we provide the total Sobol indices for the calibration points in blue (Figure~\ref{fig:data-extraction-points}) for velocity and pressure for the \textit{Classical IC} \eqref{eq:IC-mass}--\eqref{eq:IC-BJJ}. 
We observe that the boundary velocity $V^\text{top}$ has the most contribution to the velocity variance (Figure~\ref{fig:sensitivity_classical}, left) in the free-flow region and pressure field (Figure~\ref{fig:sensitivity_classical}, right) for the analyzed points in the domain.
Moreover, the exact interface location plays an important role for the velocity field, especially near the interface (Figure~\ref{fig:sensitivity_classical}, left) and influences the pressure field as well (Figure~\ref{fig:sensitivity_classical}, right). The value of the Beavers--Joseph parameter has a higher impact on the velocity near the interface than other parts of the domain. However, this parameter does not play an essential role for the pressure. The permeability $k$ significantly affects the velocity in the porous-medium domain, whereas its influence on the velocity in the free-flow region and the pressure field is small. 

\subsubsection{Generalized IC}\label{sec:sens_gic} 
\begin{figure*}[htb!]
    \centering
    \includegraphics[width=.45\linewidth]{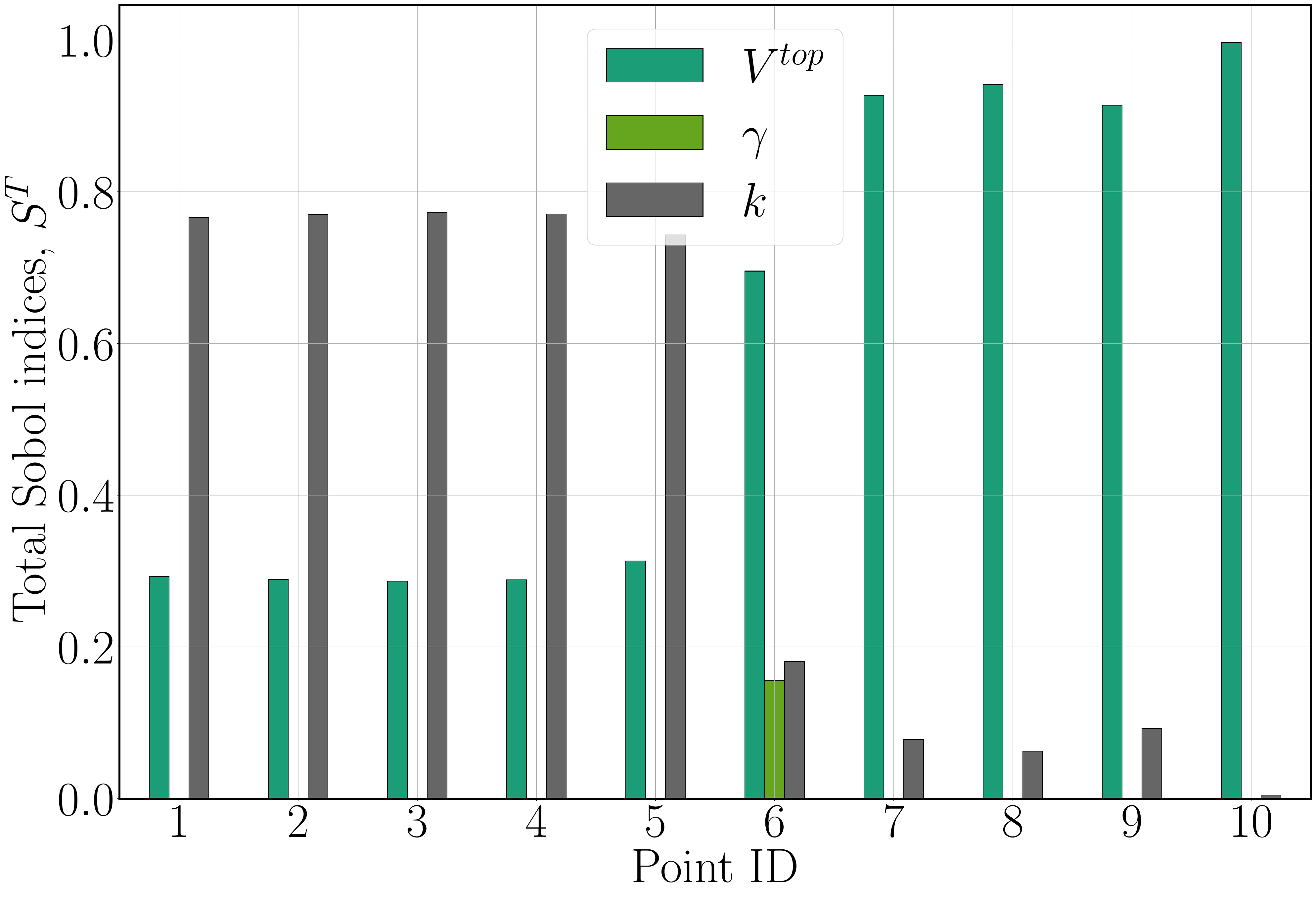} 
    \
    \includegraphics[width=.45\linewidth]{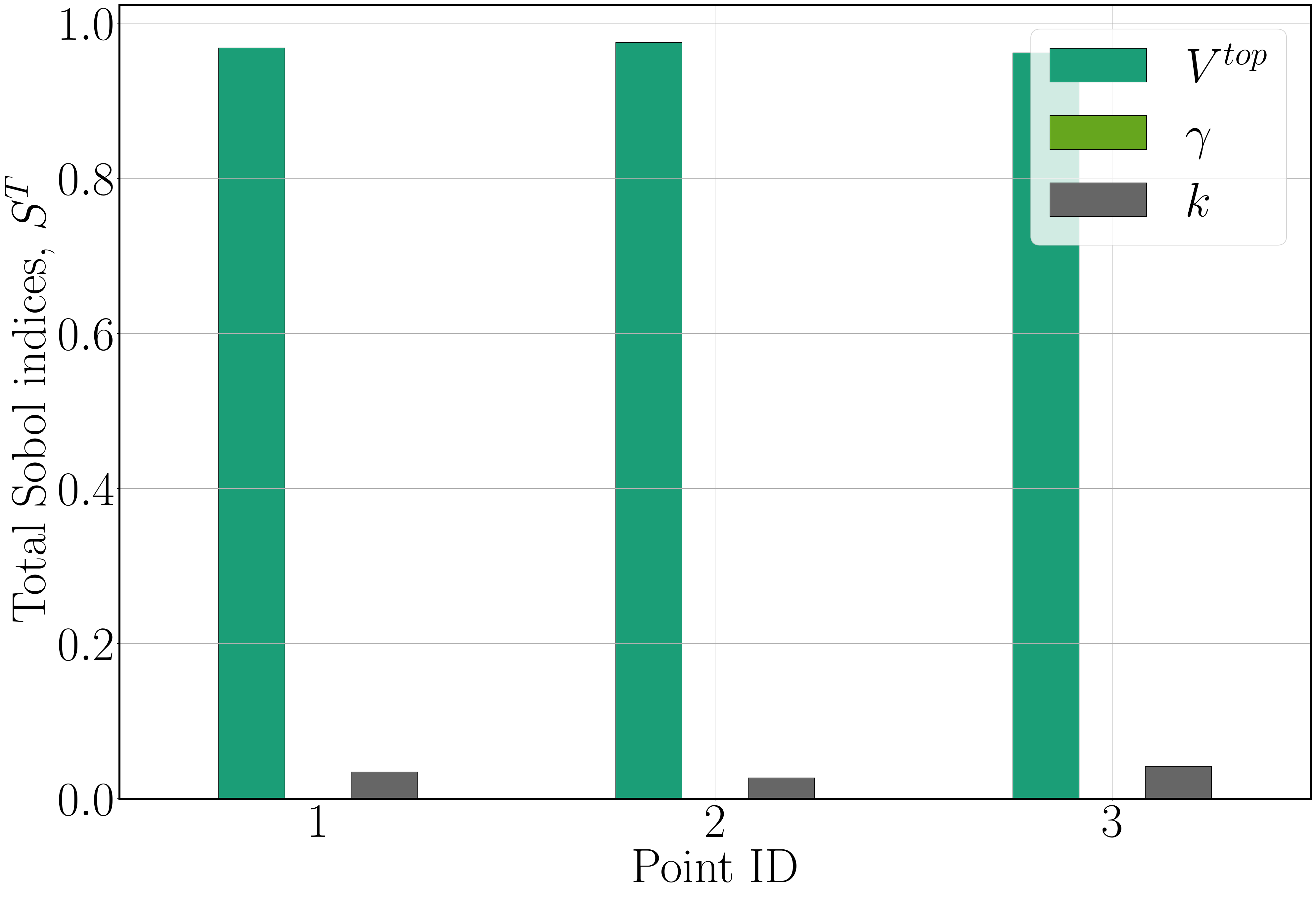} 
    \caption{Total Sobol indices of the Stokes--Darcy model with the \textit{Generalized IC} for the velocity (left) and pressure (right) for the calibration (blue) points in Figure~\ref{fig:data-extraction-points}.}
    \label{fig:sensitivity_generalized}
\end{figure*}

In  Figure~\ref{fig:sensitivity_generalized}, the total Sobol indices for velocity and pressure before calibration are presented for the selected blue points in Figure~\ref{fig:data-extraction-points}.
For the \textit{Generalized IC}, the information about the exact interface location $\gamma$ is included in the boundary layer constants $N_1^\text{bl}$ and $M_1^{1,\text{bl}}$ appearing in condition~\eqref{eq:NEW-tangential}.
Therefore, the exact position of the interface does not influence the overall system behavior in comparison to the \textit{Classical IC}.
The permeability $k$ (in the porous-medium) and the inflow velocity $V^\text{top}$ have significant impact both on the velocity (in the free-flow region) and the pressure field, as in the case of the \textit{Classical IC}.

\subsubsection{Pore-network model}\label{sec:sens_pn}
\begin{figure*}[t]
    \centering
    \includegraphics[width=.45\linewidth]{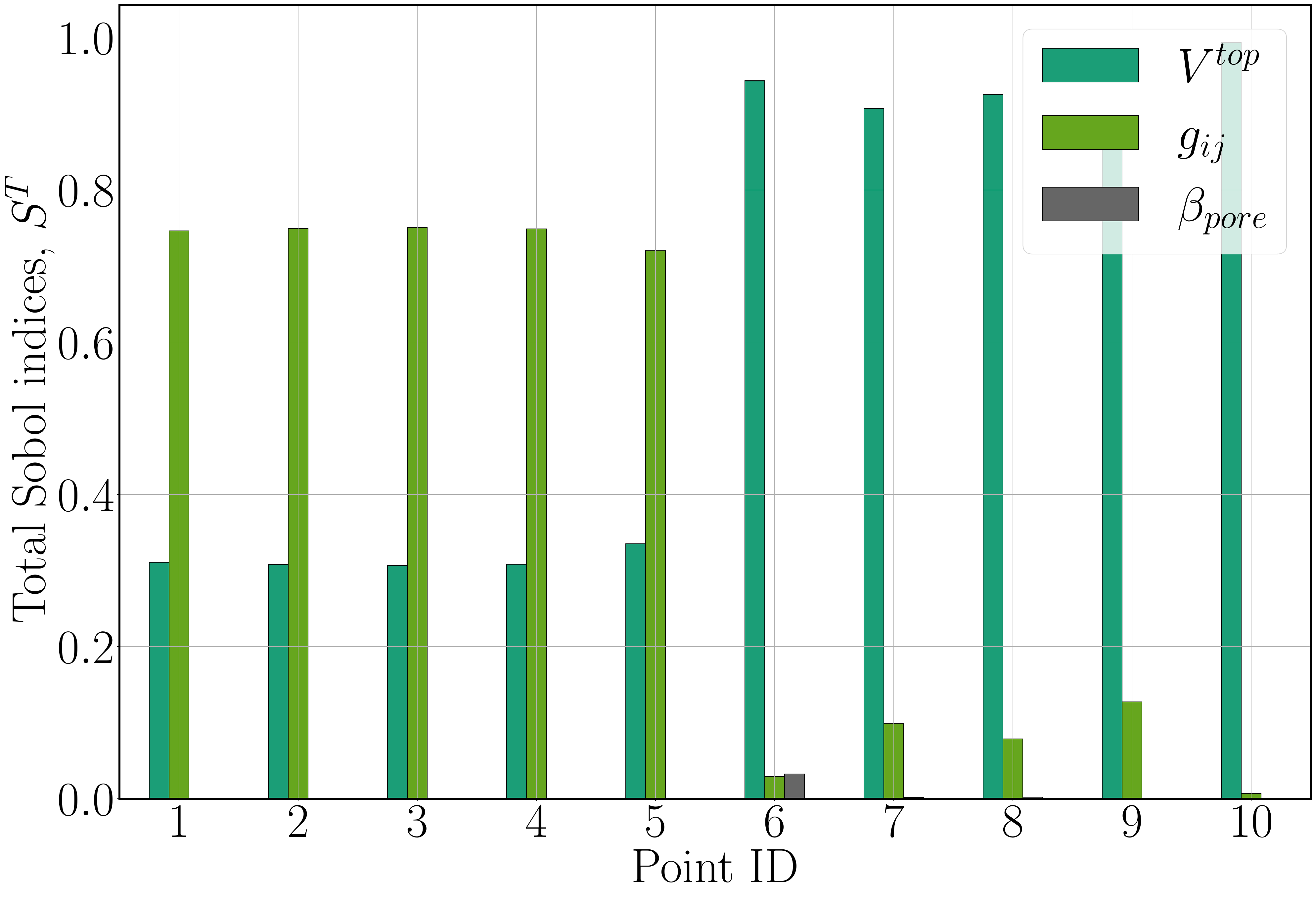} 
    \
    \includegraphics[width=.45\linewidth]{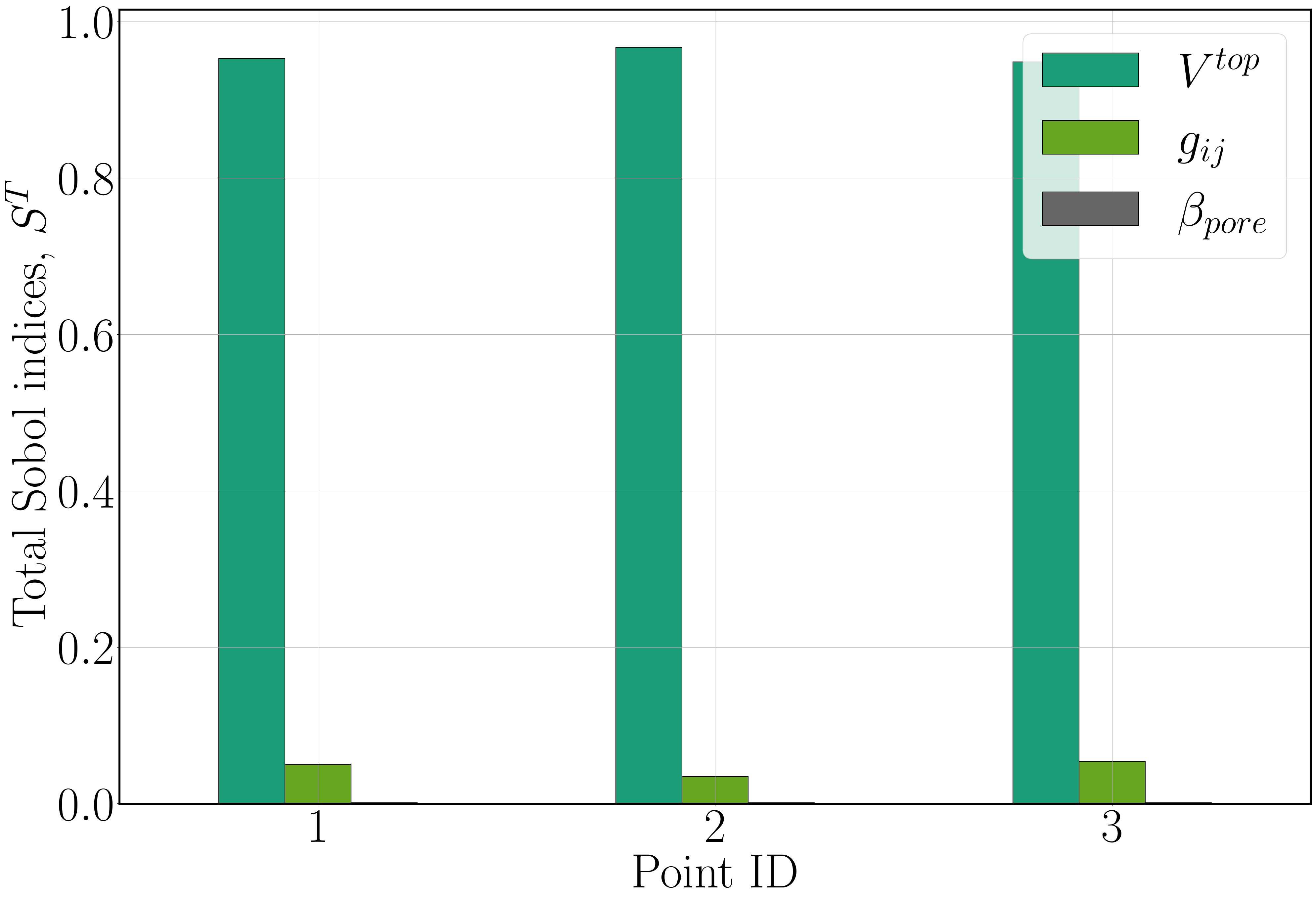} 
    \caption{Total Sobol indices of the \textit{Pore-Network} model for the velocity (left) and pressure (right) for the calibration (blue) points in Figure~\ref{fig:data-extraction-points}.}
    \label{fig:sensitivity_pnm}
\end{figure*}

Figure~\ref{fig:sensitivity_pnm} shows the total Sobol indices for velocity (left) and pressure (right) for the blue points in Figure~\ref{fig:data-extraction-points}.  
As for the REV-scale coupled models, we observe a dominant influence of $V^\mathrm{top}$ for all points. As expected, the influence of the total conductance is more prominent in the porous domain, which is comparable to the influence of permeability for the REV-scale coupled models. 
The influence of the pore-scale slip parameter $\beta_{\pore}$ shows a relatively small influence on the variability of velocity at point $6$ and is hardly visible at other locations in the free-flow region. This matter is most likely because the slip coefficient only affects the flow field in the free-flow domain $\Omega_\FF$ very locally, directly above the interface pore. 
The averaging volume used for the evaluation, however, takes into account a larger portion of the free-flow region, where the influence of $\beta_{\pore}$ is fairly small. In analogy to the REV models, $V^\mathrm{top}$ also has a dominating influence on the pressure and velocity in the free-flow region.

\subsection{Analysis of predictive abilities}\label{sec:pred_abilities}
In this section, we present the result of the analysis of the predictive ability of all three discussed conceptual models by showing their parametric posterior and the corresponding predictive distributions.
These results are generated via the surrogate-based Bayesian procedure described in Section \ref{sec:bayes}. In the calibration phase, we update the prior knowledge on the uncertain model parameters according to Section~\ref{sec:square-UQ}. We condition the responses of all analyzed models on the velocity and pressure values extracted from the pore-scale simulations that are marked as a blue points in Figure~\ref{fig:data-extraction-points}. To do so, we employ MCMC approach via \textit{emcee} Python ensemble sampling toolkit \cite{foreman2019emcee} to perform Bayesian inference in Section \ref{sec:BayesModelComparison} using BsaPCE surrogate representation in Section \ref{sec:Surrogate}. We use an Affine Invariant Ensemble Sampler (AIES) to approximate the posterior distribution. 
For more details on MCMC, we refer the reader to \cite{goodman2010ensemble}.

To accelerate this Bayesian updating step, we train each surrogate model with the simulation outcomes of 300 runs of each numerical model. The AIES-MCMC sampler is run for an ensemble of 50 Markov chains on each surrogate. We monitor the convergence of the sampler using the integrated auto-correlation time, which estimates the number of evaluations of the posterior probability density function to draw independent samples from the target density \cite{Sokal1996MonteCM}. The MCMC sampler is run until the convergence criterion of $1 \%$ for the difference in the auto-correlation time between two consequent monitoring steps is met. We retrain a new set of surrogate models in the validation stage based on the updated parameter distribution (posterior distribution) obtained after calibration. With these surrogate models, we propagate the posterior parametric uncertainty to estimate the posterior predictive distribution of models to be passed to the Bayesian metric calculation step.

As discussed in Section~\ref{sec:square-UQ}, using surrogates may introduce additional errors to the inference process. To include this error, we test the surrogate models with 150 simulation runs (test sets) which are different from the training sets. 
Comparing the surrogates' prediction with the results from the test sets, we observed a considerably low validation error between $10^{-8}$ and $10^{-11}$ for all models, indicating an acceptable prediction accuracy.
Moreover, we estimated Mean Square Error (MSE) for each surrogate model that is a good estimate of the surrogate error variance \cite{xu2015bayesian}. When evaluating the likelihood $p(\mathcal{Y}\vert M_k,\mathbf{\theta}_k)$ in \eqref{eq:BayesLikelihood}, we add a diagonal matrix $\Sigma_{\text{PCE}}$ with elements $\sigma^2_{\text{PCE},i}=MSE_i, \: i=1,2,...,N_\text{out}$ to $\Sigma$, assuming that the surrogate errors are independent and follow a normal distribution with zero mean.
Moreover, following~\cite{schoniger2015statistical}, we perturb the reference data with some additive noise to account for uncertainty associated with the BME values, the resulting Bayes factors, and posterior model weights. With this approach, we investigate the impact of other possible sources of errors on the validation metrics that are not considered in the calculations.

As mentioned in Section~\ref{sec:square-UQ}, we jointly infer the uncertain parameters with the scalar unknown parameters $\sigma^2_{vel}$ and $\sigma^2_{p}$ for each system response quantity, i.e., velocity and pressure. For these parameters, we assume uniform distributions  $ \sigma^2_{vel}\sim \mathcal{U} \left[0, 10^{-5}\right]$ and $\sigma^2_{p} \sim \mathcal{U}\left[0, 10^{-3}\right]$ as priors, for velocity and pressure, respectively.
In what follows, we present the updated (posterior) distribution of the parameters and model discrepancy errors after calibration obtained by the MCMC sampler for all three models. Afterward, a figure containing the posterior predictive of models next to each other versus the reference data is provided.

\subsubsection{Classical IC}\label{sec:pred_cic}
\begin{figure*}[h!]
    \centering
    \includegraphics[width=\linewidth]{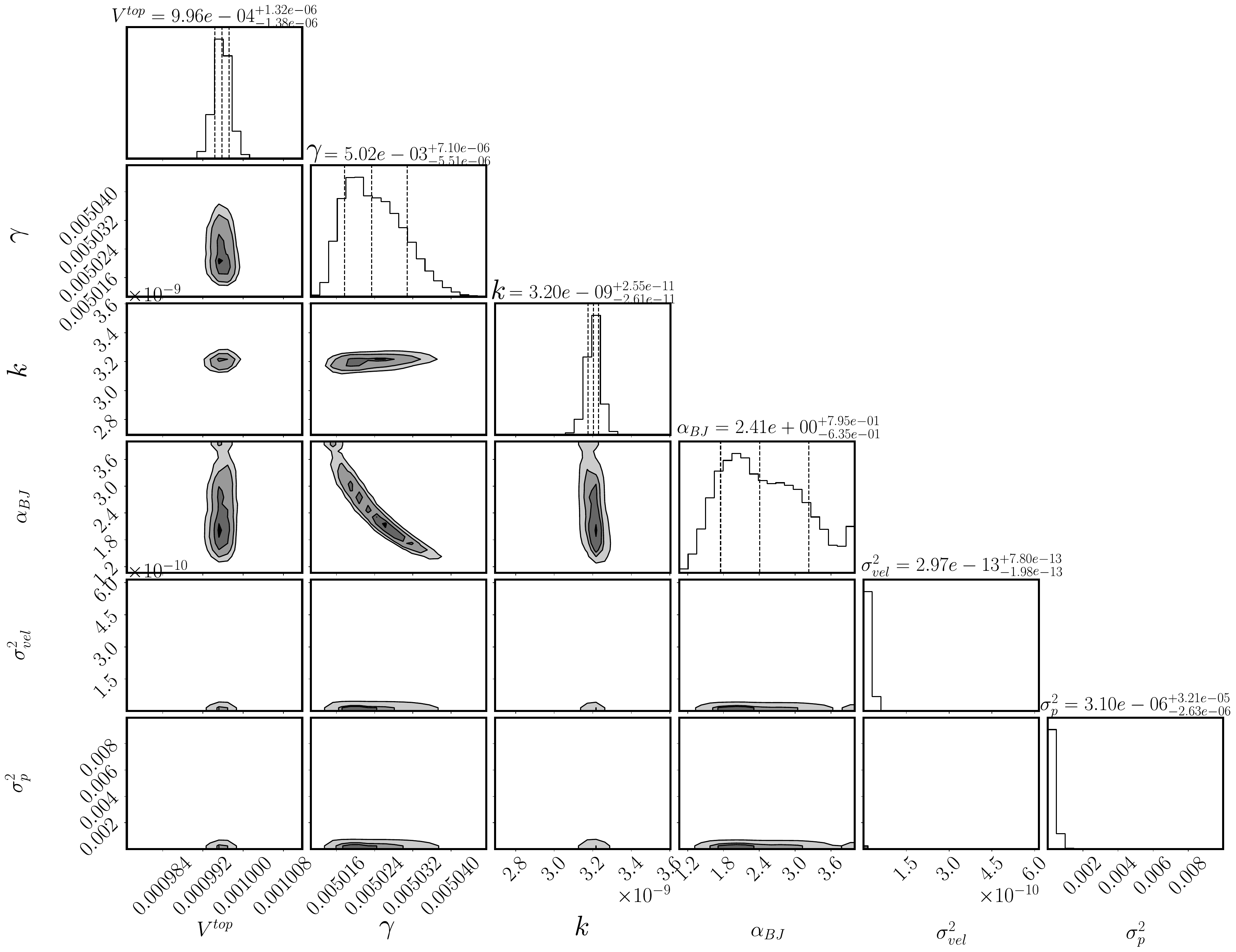}
    \caption{Posterior parameter distribution of the Stokes--Darcy model with the \textit{Classical IC} after calibration to the reference data from the pore-scale model.}
    \label{fig:postrior_cic}
\end{figure*}

Figure~\ref{fig:postrior_cic} presents the posterior distribution obtained via the  Bayesian inference using the calibration (blue) points in Figure~\ref{fig:data-extraction-points}. The 50 percent quantiles, alongside the 15 and 85 percent quantiles, are displayed on top of the histograms shown in the diagonal plots. 
Most posterior distributions of the parameters follow a Gaussian distribution. However, the distribution of the interface location $\gamma$ and the Beavers--Joseph slip coefficient  $\alpha_{BJ}$ exhibits a long right tail.
Moreover, a slight correlation between $\gamma$ and $\alpha_\BJ$ is observed. 

\subsubsection{Generalized IC}\label{sec:pred_gic} 
\begin{figure*}[h!]
    \centering
    \includegraphics[width=.8\linewidth]{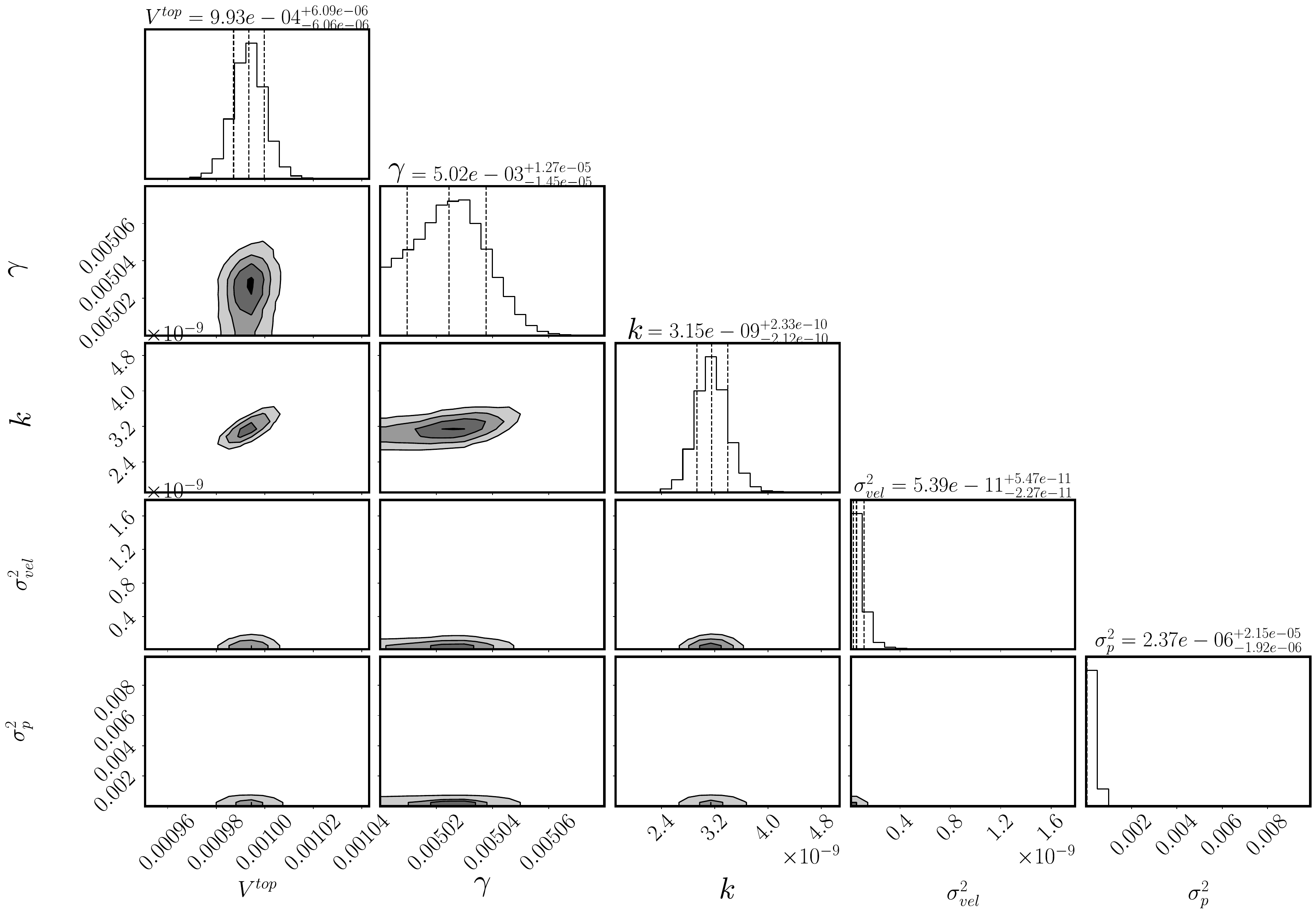}
    \caption{Posterior parameter distribution of the Stokes--Darcy model with the \textit{Generalized IC} after calibration to the reference data from the pore-scale model.}
    \label{fig:postrior_gic}
\end{figure*}

Similar to the procedure described above, the surrogate-based Bayesian calibration offers insight into the posterior distributions of modeling parameters for the Stokes--Darcy model with the \textit{Generalized IC} (Figure~\ref{fig:postrior_gic}). 
As opposed to the \textit{Classical IC}, the interface location $\gamma$ for this coupling condition shows a slightly wider distribution.
This observation indicates that the exact position of interface does not influence the overall system behavior in comparison to the \textit{Classical IC}. 

\begin{figure*}[t!]
    \centering
    \includegraphics[width=.8\linewidth]{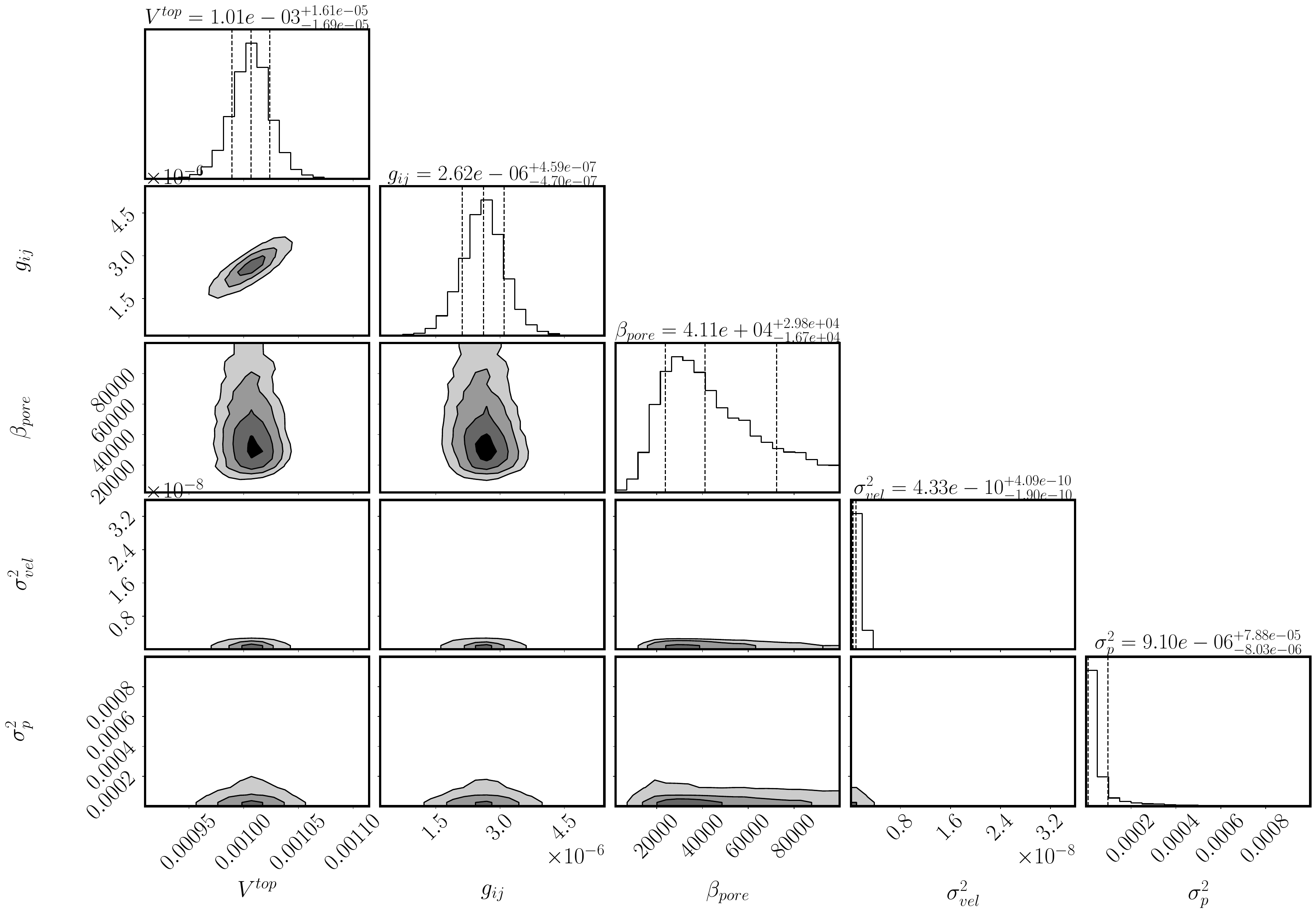}
    \caption{Posterior parameter distribution of the \textit{Pore-Network} model after calibration to the reference data from the pore-scale model.}
    \label{fig:postrior_pnm}
\end{figure*}

\subsubsection{Pore-network model}\label{sec:pred_pn} 
For the \textit{Pore-Network} model, we also have used the calibration (blue) points (Figure~\ref{fig:data-extraction-points}) to perform surrogate-based Bayesian inference. Figure~\ref{fig:postrior_pnm} illustrates the posterior parameter distribution of the \textit{Pore-Network} model. 
The distribution of $\beta_{\pore}$ covers a wider range. This issue can be attributed to insensitivity of the model results to this parameter, as presented by the total Sobol indices in Figure~\ref{fig:sensitivity_pnm}. 

\begin{figure*}[h!]
    \centering
    \includegraphics[width=.75\linewidth]{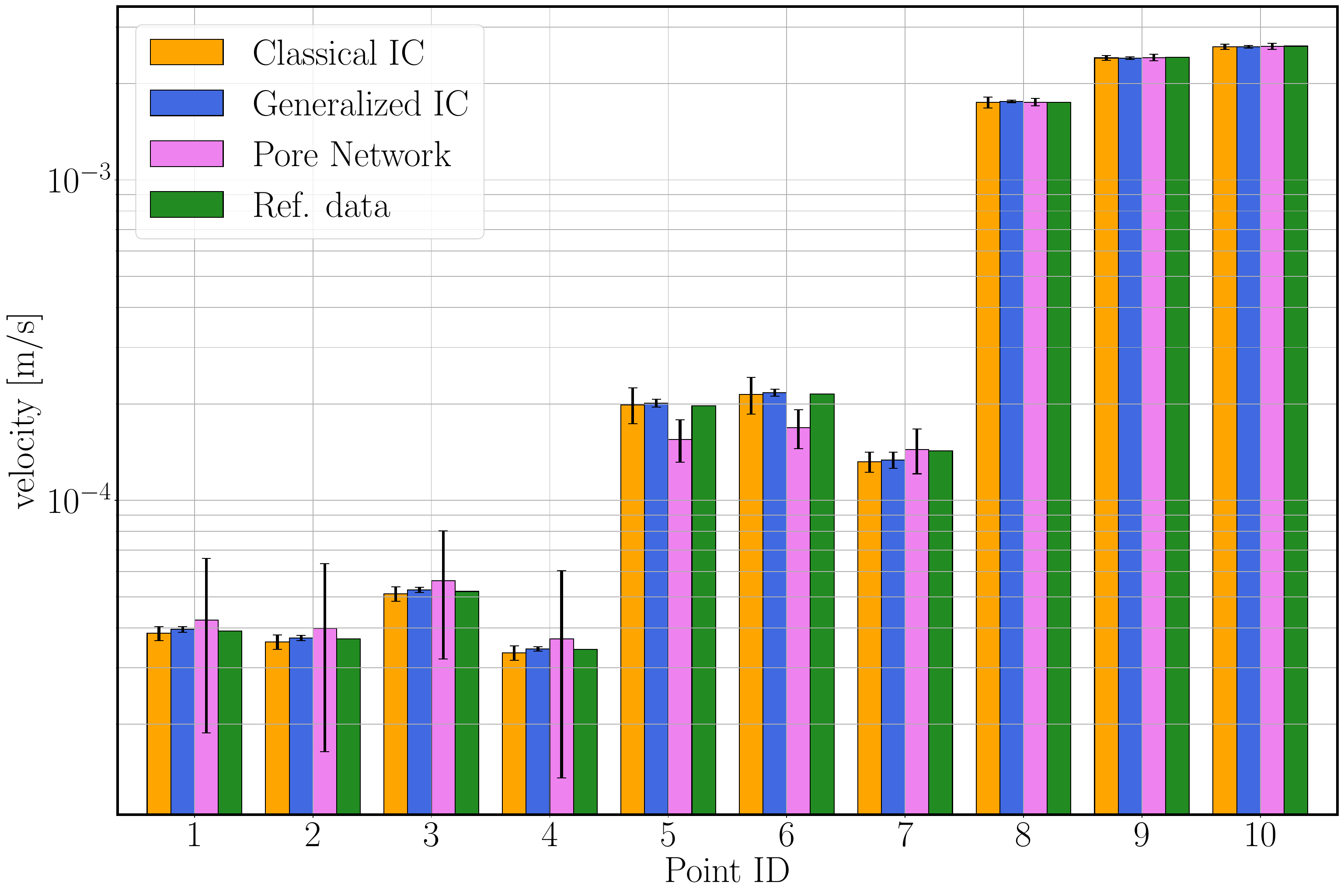} 
    \caption{The velocity predictions of all models in the validation step against the reference data from the pore-scale model.}
  \label{fig:pred_vel}
\end{figure*}

\begin{figure*}[h!]
    \centering
    \includegraphics[width=.8\linewidth]{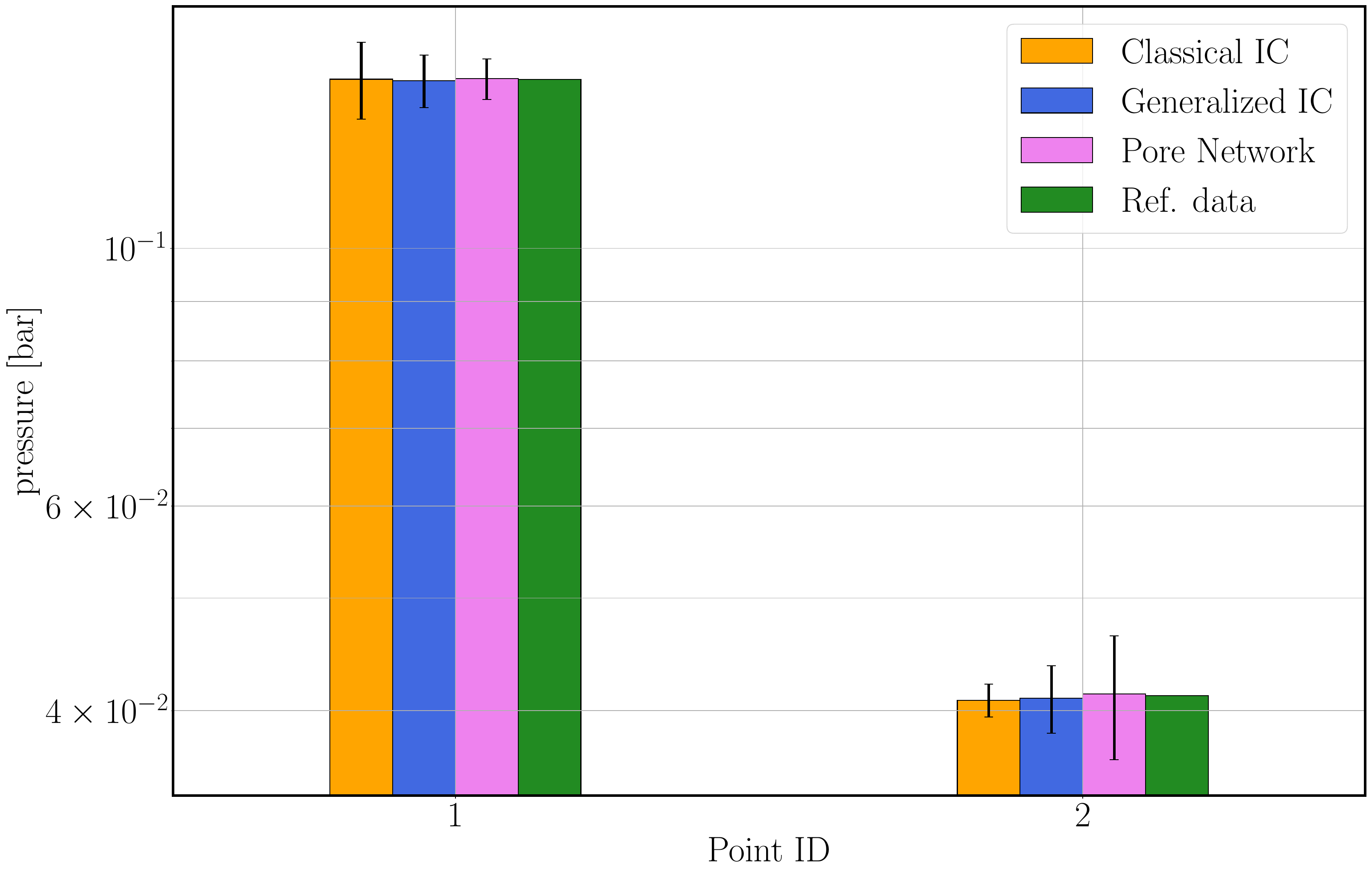} 
    \caption{The pressure predictions of all models in the validation step against the reference data from the pore-scale model.}
  \label{fig:pred_p}
\end{figure*}

To obtain the models' posterior predictive distributions, we need to propagate the posterior parametric uncertainty presented so far through the models. The result offers a possibility of analyzing how post-calibration uncertainty affects the SRQs. 
To perform the post-calibration uncertainty propagation, we have trained a new surrogate for each competing model using new training sample points drawn from the posterior parameter distribution.
For better visual comparison, we plot the posterior predictive of models next to each other.
Figures \ref{fig:pred_vel} and \ref{fig:pred_p} illustrate the mean and standard deviations of the model predictive distributions in a bar chart for the velocity and pressure response quantities, respectively. 

In particular, Figure~\ref{fig:pred_vel} reveals that all analyzed models provide accurate predictions at the points located in the deeper part of the porous medium (1 to 4). However, the predictions at the points near the interface (5 to 8) suggest that the Stokes--Darcy model with \textit{Classical IC} and \textit{Generalized IC} provide more accurate predictions than the \textit{Pore-Network} model. The REV-scale model with  \textit{Generalized IC} shows less uncertainty, i.e., lower standard deviation, in its prediction at the vicinity of the interface between the porous medium and the free-flow.
Moreover, Figure~\ref{fig:pred_p} confirms that all models are able to provide accurate pressure values.

\subsection{Model comparison}\label{sec:model_comparison}
We perform model comparison employing the so-called posterior model weights according to the Bayesian approach explained in Section~\ref{sec:BayesModelComparison}. Such an analysis offers an aggregated comparison of a model's outputs to the validation set of reference data from the pore-scale model that are marked in red in Figure~\ref{fig:data-extraction-points}. For model comparison analysis, we use the newly constructed surrogate representation during the validation stage to compute the BME values in \eqref{eq:BME}. These values are required to calculate the posterior model weights~\eqref{eq:BayesBMA} and the Bayes factors in \eqref{eq:BayesFactor}.
Additionally, use of the advanced surrogate representation provides a possibility to assess uncertainty of the BME values and the corresponding model weights.
Table~\ref{tab:post_model_weights} presents a detailed statistical summary of the model weights and provides a ranking. It also reports the information regarding the post-calibration uncertainty with help of the deviation regarding 25$\%$ and 75$\%$ percentiles.

\begin{table}[!ht]
\renewcommand*{\arraystretch}{1.4}
\caption{The statistical summary of posterior model weights after validation.}
\centering
\begin{tabular}{l c c}
\hline
 Model & Model weights & Rank  \\
\hline
    Classical IC & $0.003^{+0.002}_{-0.001}$ & 2 \\
    Generalized IC & $0.997^{+0.001}_{-0.002}$ & 1 \\
    Pore-network & $0.000^{+0.000}_{-0.000}$ & 3 \\
\hline
\end{tabular}
\label{tab:post_model_weights}
\end{table}
The expected model weights under noisy pore-scale data assumption convey a relatively clear model ranking in favor of \textit{Generalized IC}, with \textit{Classical IC} as second and the \textit{Pore-Network} model ranking last.
It is worth mentioning that the model weights close to zero for the \textit{Classical IC} and \textit{Pore-Network} model can be attributed to the high prediction uncertainty of these models. This fact is represented by the error bars in Figure~\ref{fig:pred_vel}. Moreover, a considerable mismatch can be detected between the expected velocity prediction of the \textit{Pore-Network} model and the reference data at validation points 5 and 6.
As for the \textit{Classical IC}, the velocity prediction uncertainty is higher than that of the \textit{Generalized IC}. This difference is mainly for the points at the vicinity of the interface and in the free-flow region.

\begin{figure*}[t]
    \centering
    \includegraphics[width=.75\linewidth]{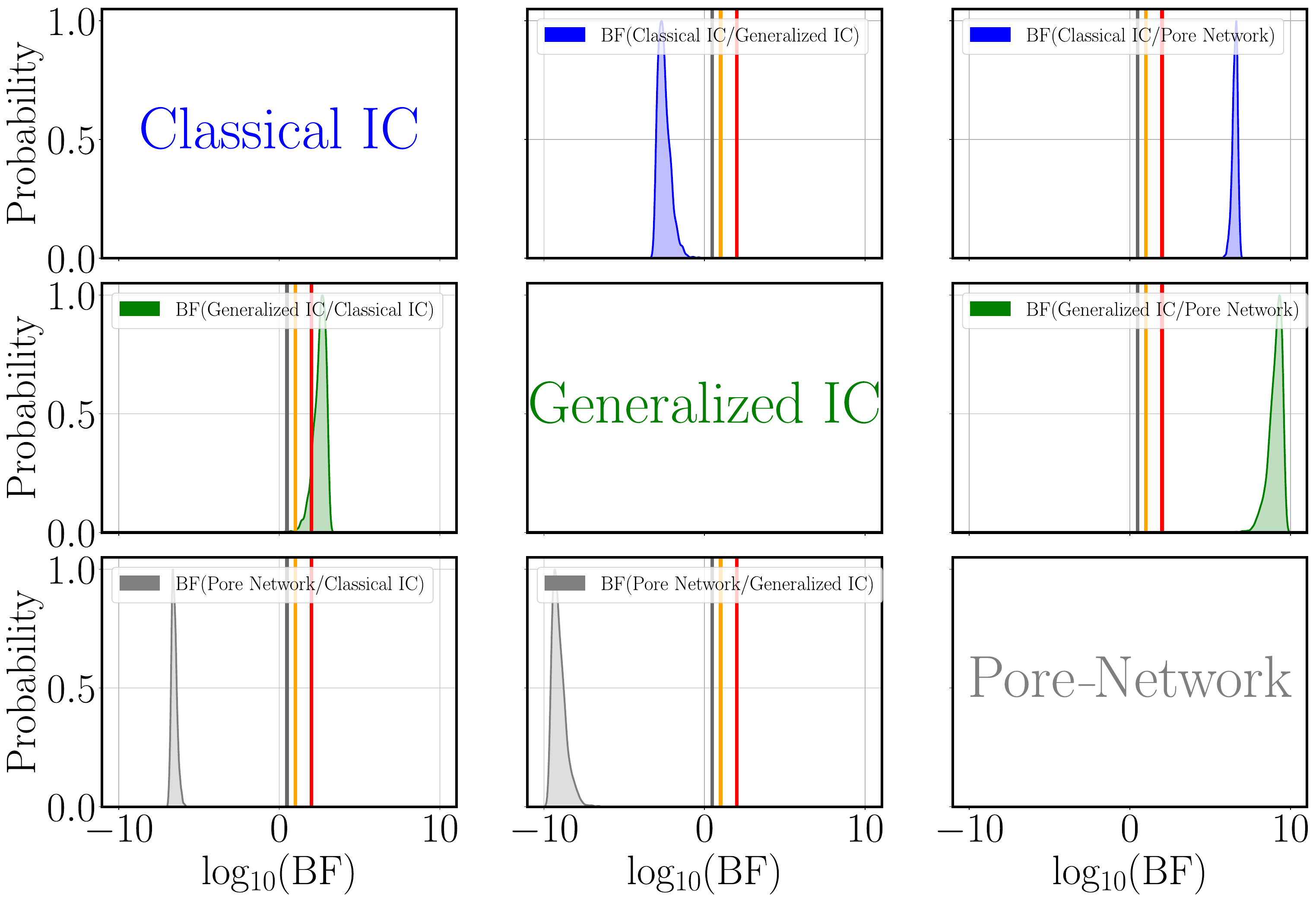} 
    \caption{Distributions of log$_{10}$ (Bayes Factor) for the pairwise comparison of competing models based on the validation scenario.}
    \label{fig:BayesFactor_plot}
\end{figure*}

Assessments of confidence in model ranking have been investigated by means of Bayes Factor~\eqref{eq:BayesFactor} for pairwise comparison of models based on the validation scenario. In the introduced uncertainty-aware Bayesian validation framework, Bayes Factors provide an objective measure of significance that quantifies the evidence in favor of one model's superiority against another.
Figure~\ref{fig:BayesFactor_plot} presents the probability density functions of log$_{10}$(BF) over all perturbed velocity and pressure data sets in a three-by-three matrix. Here, we compute three Bayes Factors for each model against its counterpart. The significant levels in a log$_{10}$-scale, introduced in \cite{jeffreys1961theory} are marked with the vertical lines.
Gray lines represent equally strong evidence for both models. Orange and red lines indicate thresholds for strong and decisive evidence in favor of one model against the other, respectively. 

The first plot in the second row in~Figure~\ref{fig:BayesFactor_plot}, e.g., shows the distribution of log$_{10}$(BF) in favor of \textit{Generalized IC} against \textit{Classical IC}. This plot reveals that for most of the perturbed data sets, the Bayes factor is in the region where a decisive evidence (log$_{10}$(BF) greater than two) exists in favor of \textit{Generalized IC} to outperform \textit{Classical IC}.
Similarly, in all the analyzed cases (perturbed data sets), \textit{Classical IC} could be clearly favored against \textit{Pore-Network} model based on the decisive evidence (the plot in the first row, the last column).
Moreover, the distribution in the second row, third column of Figure~\ref{fig:BayesFactor_plot} reveals that the Bayes factor distribution of \textit{Generalized IC} against \textit{Pore-Network} model proves a decisive evidence in favor of \textit{Generalized IC}.

\begin{figure*}[t]
    \centering
    \includegraphics[width=0.75\linewidth]{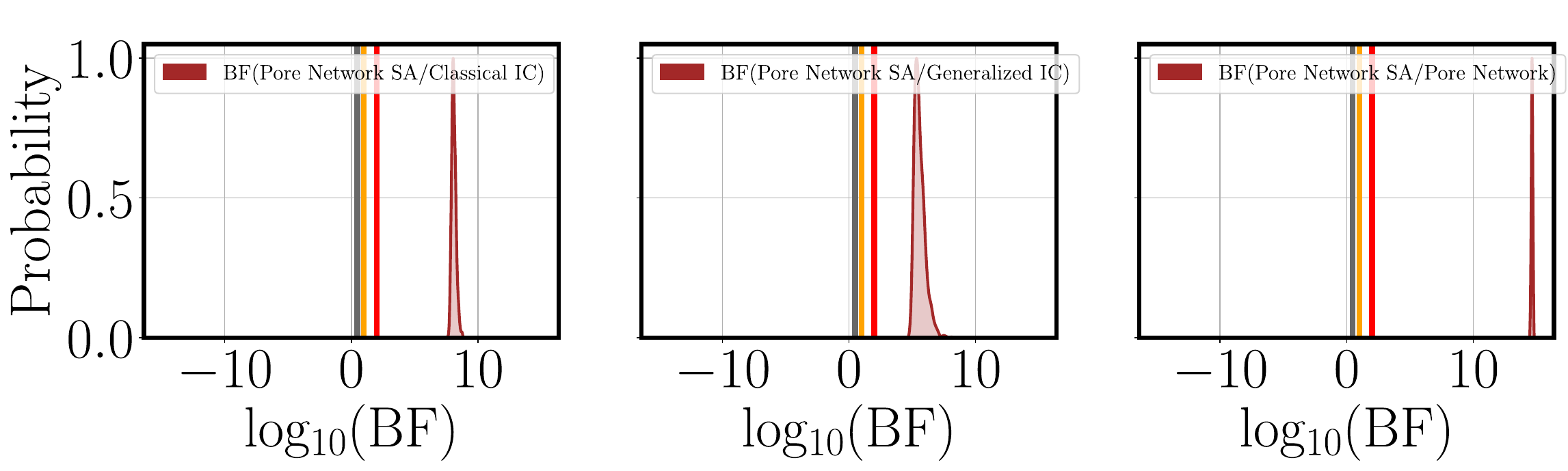} 
    \caption{Distributions of log$_{10}$ (Bayes factor) of the \textit{Pore-Network} model, with the surface averaging against competing models based on the validation scenario.}
    \label{fig:BayesFactor_plot_NA}
\end{figure*}

The results presented so far are based on a comparison of the SRQs with averaged SRQs of the fully resolved Stokes simulation, as the Stokes--Darcy model with \textit{Classical IC} and \textit{Generalized IC} could offer a prediction on the REV scale only. 
However, one could directly compare the \textit{Pore-Network} model to the reference data at the pore scale without performing volume averaging by calculating the surface-averaged pore-scale velocity at the pore-throat cross-sections. We denote the \textit{Pore-Network} model with pore-throat surface averaging model as the \textit{Pore-Network SA} model and its Bayes factors distribution is shown in Figure~\ref{fig:BayesFactor_plot_NA}. The velocities of \textit{Pore-Network SA} model are not defined within the pore bodies but only at the pore throats, which explains why the results of Figure~\ref{fig:BayesFactor_plot_NA} show stronger evidence in favor of the \textit{Pore-Network SA} model compared to the other concepts. Therefore, the \textit{Pore-Network SA} model avoiding additional averaging steps is a suitable approach when detailed pore-scale information is considered. Alternatively,  the Stokes--Darcy model with \textit{Generalized IC} adequately represents the underlying physical processes once the REV-scale information is available only.

In addition to the setup presented in Section~\ref{sec:square-BC}, we also analyzed two other cases. Firstly, we considered a setup with the same geometrical configuration, but the inlet boundary was located at the left domain edge in the free-flow region with an opening of 1.5 mm from the top. This setup induces a flow profile parallel to the interface. 
Comparing \textit{Classical IC} with \textit{Generalized IC}, we witnessed no substantial evidence in favor of any model. This observation is in line with the results from~\cite{Eggenweiler_Rybak_MMS20}, where the authors showed that the Stokes--Darcy problem with \textit{Classical IC} and \textit{Generalized IC} provides similar simulation results for parallel flows to the porous layer. 
The second additional setup is based on the same flow models and boundary conditions as presented in Section~\ref{sec:square-BC}, however, the solid inclusions are circular. We compared the Stokes--Darcy model with \textit{Classical IC} and \textit{Generalized IC} against the reference data. The model comparison with Bayes factor suggests strong evidence in favor of the \textit{Generalized IC}, as expected and similar to the rectangular inclusions.

\section{Summary and conclusions}
\label{sec:conclusions}
We have proposed a surrogate-assisted uncertainty-aware Bayesian validation framework and applied it to a benchmark study that addresses not only these parametric uncertainties, but also conceptual modeling uncertainties due to different formulations of physical models. 
To do so, we have considered the Stokes equations coupled to different models for the porous-medium compartment and corresponding coupling strategies: the standard REV-scale model using Darcy’s law with classical or generalized interface conditions as well as the pore-network model. 
The advantage of employing a surrogate modeling technique is that one can perform a sensitivity analysis without additional costs. This analysis is achieved using the so-called Sobol indices that are derived analytically from the expansion coefficients.
The application of the presented surrogate-assisted Bayesian uncertainty-aware framework is not limited to the models considered in this manuscript, but can be applied to many other applications.

Applying the suggested Bayesian validation framework, we have observed that there are matches between the predictions related to the considered models and the reference data for the points in the deeper part of the porous medium for all coupled models. However, we have found differences in the predictive capabilities of the models in the vicinity of the interface and in the free-flow region.
Moreover, we have propagated the post-calibration parametric uncertainty through each analyzed model to validate the different models against reference data that have not been used during the calibration phase. This uncertainty-aware Bayesian validation procedure has confirmed that the averaged pore-network model has the most difficulties representing the underlying physical process correctly. This issue is most likely due to the averaging approach used for the pore-network model, where velocities have to be calculated and interpolated from fluxes that are only given within pore throats. Moreover, addressing the differences in the predictions of the considered modeling concepts, we have performed a Bayesian model comparison. This comparison reveals that the Stokes--Darcy model with the generalized interface conditions represents processes on the REV scale best compared to the classical interface conditions and the correspondingly upscaled pore-network model. The pore-network model outperforms both Stokes--Darcy models with classical and generalized interface conditions only 
if the detailed pore-scale information is considered.
We have also investigated two other cases: one with the opening boundary condition on the left side and another with circular inclusions in the porous medium. The analysis of the former setting, which induces parallel flow to the interface, reveals that the Stokes--Darcy models with the classical and the generalized interface conditions provide similar results. This observation was expected for flows parallel to the fluid--porous interface. 
The findings of the analysis for the setup with circular inclusions confirm that there is decisive evidence in favor of the generalized interface condition being superior to the classical interface.
Concluding, we have observed that the suggested surrogate-assisted uncertainty-aware Bayesian validation framework helps to gain insight into underlying physical processes at considerably low computational costs.

\bmhead{Acknowledgments}
The work is funded by the Deutsche Forschungsgemeinschaft (DFG, German Research Foundation) -- Project Number 327154368 -- SFB 1313.

\bmhead{Statements and Declarations}
\begin{itemize}
    \item \textbf{Funding} Open Access funding enabled and organized by Projekt DEAL.
    \item \textbf{Competing interest} The authors have no competing interests to declare that are relevant to the content of this article.
    \item \textbf{Data availability} The Bayesian framework and the models' source codes, as well as the reference data used in this study, are available at \url{https://git.iws.uni-stuttgart.de/dumux-pub/mohammadi2022a}.
\end{itemize}



\end{document}